\documentclass[prl,twocolumn,aps,showpacs]{revtex4}

\usepackage{graphicx}
\usepackage{bm}
\usepackage{amsmath}
\usepackage{bbm}

\newcommand{\ket}[1]{\left| #1 \right>} 
\newcommand{\bra}[1]{\left< #1 \right|} 

\begin{document}

\title{Non-Abelian Topological Order from \\ Quantum Organization in Indistinguishable Groups}

\author{Bel\'{e}n Paredes}

\affiliation{Instituto de F\'{i}sica Te\'{o}rica CSIC/UAM \\C/Nicol\'{a}s Cabrera, 13-15
Cantoblanco, 28049 Madrid, Spain}

\date{\today}

\begin{abstract} 
I propose that non-Abelian topological order can emerge from the organization of quantum particles into identical indistinguishable copies of the same quantum many-body state. Quantum indistinguishability (symmetrization) of the collectivities leads to topological degeneracy in the subspace of elementary excitations, giving rise to non-Abelian braiding statistics. The non-Abelian hidden order of a symmetrized structure is manifested in its entanglement properties, and the corresponding non-Abelian fusion and braiding rules can be derived by analyzing the set of symmetrized states  on a surface with non-trivial topology like a torus. To illustrate the emergence of non-Abelian statistics from symmetrization, I consider the case of two identical copies of the toric code model. The resulting model is shown to be non-Abelian, exhibiting two types (charge and flux) of quasiparticles with non trivial fusion channels. The symmetrization construction I present here constitutes a framework for the generation of non-Abelian models from known Abelian ones.
\end{abstract}


\maketitle

\section{INTRODUCTION}

Quantum Hall systems \cite{FQHE,Laughlin,Wen90,Wen91} taught us that we should not regard quantum statistics as the result of permuting variables in a many-body wave function, but rather as the one of exchanging particles along a physical path \cite{Wilczek}. Within this definition, quantum statistics in two dimensions is a representation of the Braid group \cite{Khare}, where a braid describes the physical exchange of the world-lines of particles. One-dimensional Abelian representations of the Braid group correspond to fractional statistics, with statistical phase interpolating between the one of bosons and fermions. Remarkably, non-Abelian representations open the possibility for the existence of quasiparticles with non-Abelian braiding statistics 
\cite{MooreRead91,Wen91_NA,Nayak2008,Stern2008,Stern2010}. The concept of non-Abelian statistics is profoundly unintuitive, since it implies that the exchange of identical quasiparticles produces a global change in the underlying vacuum, converting it into a topologically different one. The space of topologically distinct vacua's, connected through braidings of identical quasiparticles, 
exhibits a highly non-local structure in which quantum information can be stored safely, 
immune to local interactions with the environment \cite{Nayak2008,Kitaev2003,Preskill2002}.

\begin{figure}[t]
\includegraphics[width=\linewidth]{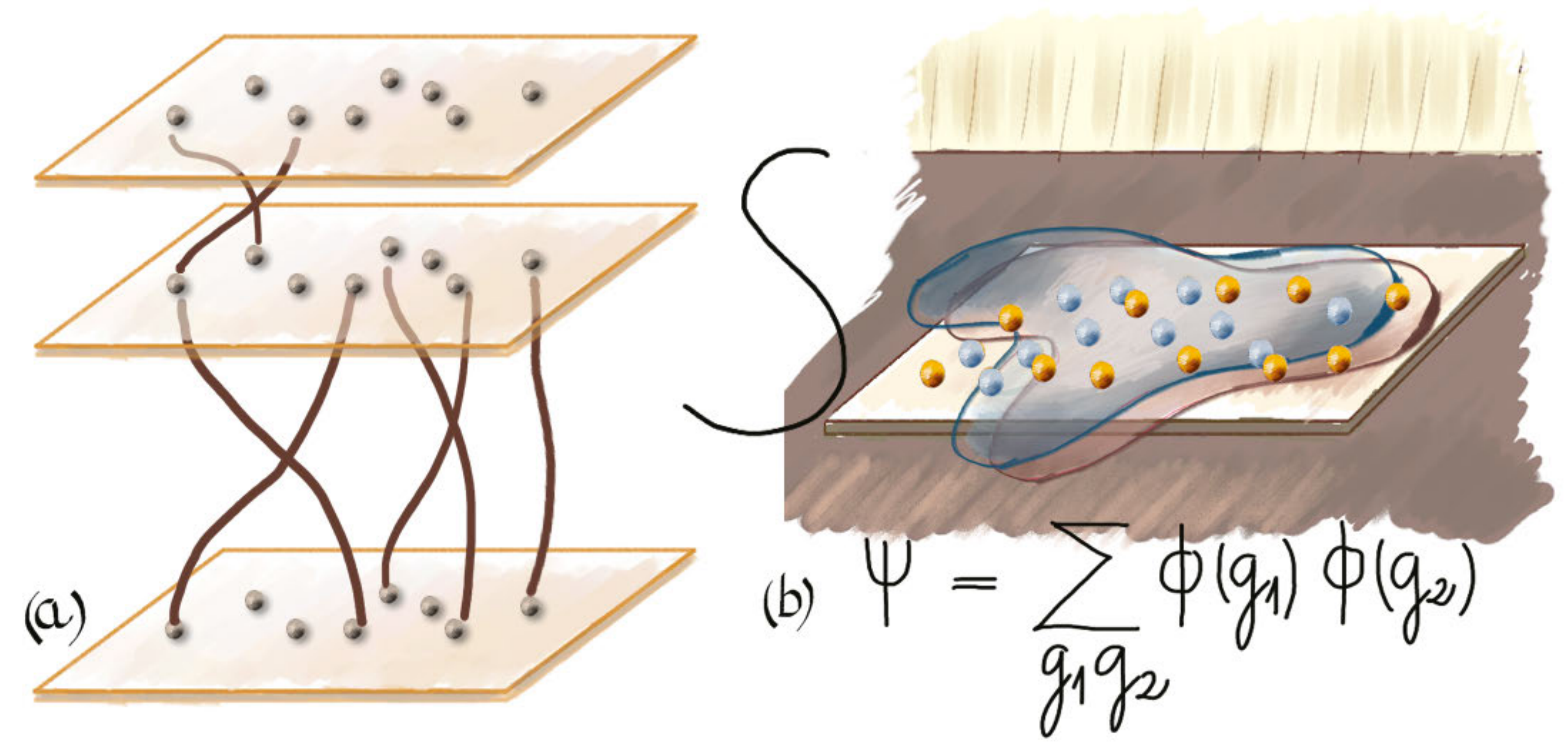}
\caption{\textbf{Non-Abelian anyons and symmetrization}. (\textbf{a}) Manifold of topologically distinct many-body states connected through braiding of non-Abelian quasiparticles. Quantum organization in identical copies of a certain many-body state (\textbf{b}) is proposed to lead to non-Abelian anyons.
}
\label{NAAnyons}
\end{figure}

Non-Abelian anyons have been predicted to occur both in fractional quantum Hall systems \cite{MooreRead91,Wen91_NA}, as excitations of the so called Pfaffian state \cite{PfaffianExp,Dolev2008,Radu2008}, and as Majorana fermions attached to vortices in superconductors with 
$p$-wave pairing \cite{Ivanov2001,Kane2008}. Recently, evidence of such Majorana fermions has been found in semiconductor nanowires coupled to superconductors \cite{MajoranaKouwenhoven}. Towards the fundamental understanding of non-Abelian topological order, the discovery of exactly solvable models exhibiting non-Abelian phases has played a crucial role \cite{Kitaev2003,WenBook,Levin2005,Freedman2004,Freedman2005,Kitaev2006,Fendley2008,Troyer2008,BombinColorCodes,Bombin2008,Fidkowski2009,Velenich2010,Wen99}.
In this direction, the spin lattice models developed by Kitaev \cite{Kitaev2003,Kitaev2006}, together with quantum loop models \cite{Freedman2004,Freedman2005,Fendley2008,Troyer2008,Fidkowski2009,Velenich2010} and string-net models \cite{WenBook,Levin2005,BombinColorCodes},
constitute eminent examples. Non-Abelian anyons have been also proposed to appear when introducing non-trivial scalar products \cite{Fendley2008} or twists \cite{Bombin2010} in quantum loop models.
Remarkable insight into the underlying properties of non-Abelian topological order has been gained by Wen's theory 
\cite{WenBook,Levin2005}, 
in which both Abelian and non-Abelian states are proposed to emerge from the condensation of extended objects dubbed string-nets \cite{Levin2005}.
String-net condensation reveals that the mathematical framework underlying topological order is tensor category. Tensor network representations for the ground states of string-net models have been developed in \cite{Buerschaper2009,Pfeifer2010}. 
But, as much as this endeavor has been fruitful, it is also far from being complete. 
The challenge remains to find descriptions of non-Abelian states that allow us to characterize them in an intuitive way, identifying the features of their underlying Hamiltonians and paving the way towards their experimental realization.

\begin{figure}[t]
\includegraphics[width=\linewidth]{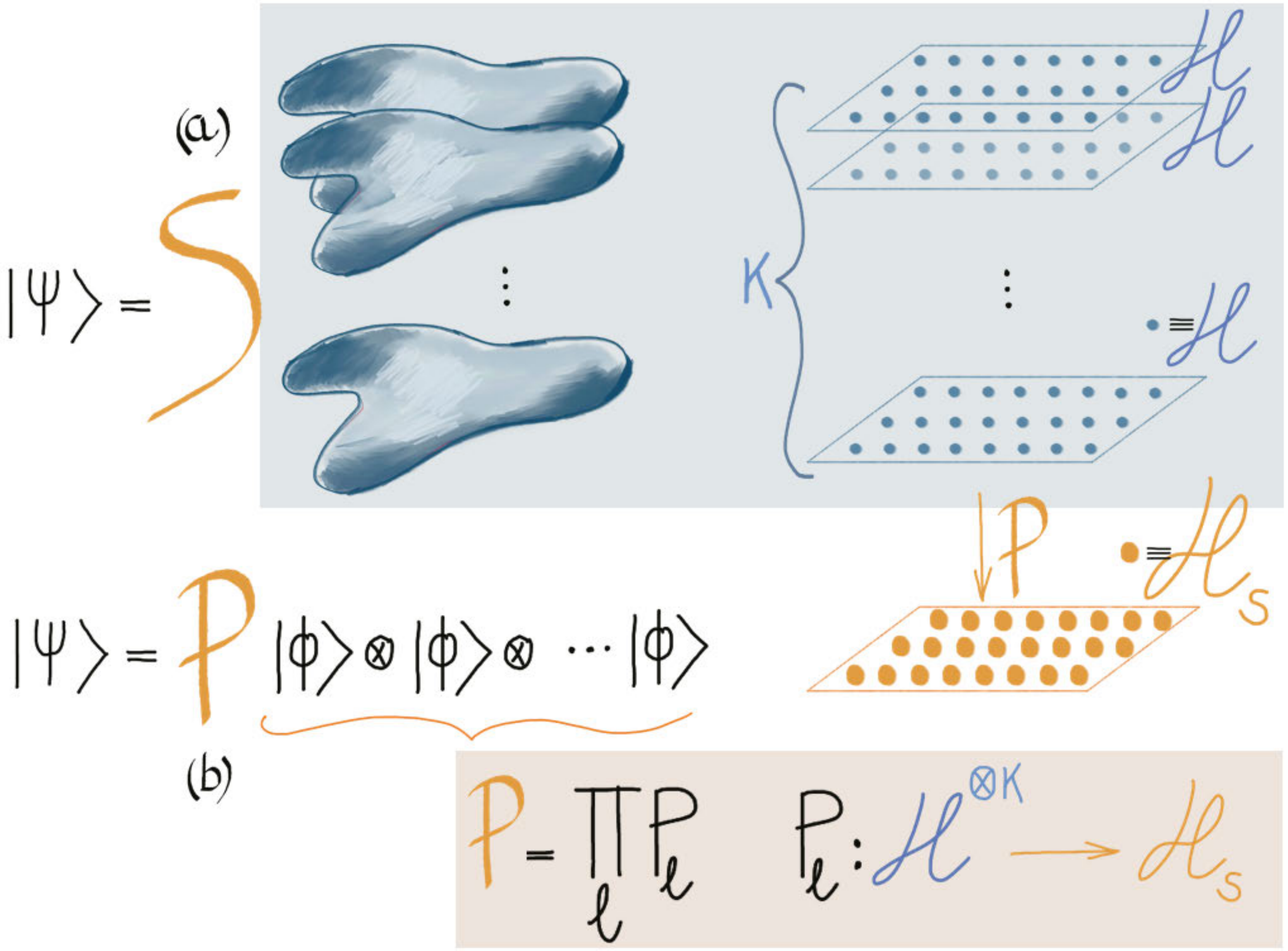}
\caption{\textbf{Quantum organization in indistinguishable copies}. Identical copies of the same quantum many-body state are merged through the symmetrization $\mathcal{S}$ (\textbf{a}). For a lattice system of local degrees of freedom described by the Hilbert space $\mathcal{H}$, the symmetrization is realized by the projector $\mathbf{P}$ (\textbf{b}), a product of local projectors which map the tensor product of the identical Hilbert spaces onto the symmetric Hilbert space $\mathcal{H}^S$.
}
\label{Symmetrization}
\end{figure}	


Here, I propose that non-Abelian topological order can arise from the organization of quantum degrees of freedom into indistinguishable copies of the same quantum many-body state
(Fig.\,\ref{NAAnyons}).  The non-commutative algebra characterizing non-Abelian braiding statistics originates from the symmetrization of the identical Abelian algebras defining the Abelian statistics of the copies.
This approach aims at providing an intuitive picture for the physical mechanism leading to the formation of non-Abelian anyons, giving a simple description of the pattern of many-body entanglement underlying non-Abelian topological order. Such pattern
is inspired by the form of the wave function describing non-Abelian fractional quantum Hall liquids \cite{MooreRead91}, which can be constructed from copies of Abelian quantum Hall states 
\cite{Wen99,WenProjectiveConstruction,Paredes2012a}. This work is motivated by my previous work on the emergence of non-Abelian braiding properties from merging of identical quantum loop condensates 
\cite{Paredes2012b}. 

The theory is presented in three stages: {\em intuition}, {\em conjecture} and {\em illustration}.
{\em First}, I explain the intuition behind the general idea, arguing how the symmetrization of identical quantum many-body states leads to the origin of non-Abelian quasiparticles. The reasoning I follow is simple. 
The freedom to assign quasiparticles to the identical copies gives rise to a set of degenerate states that are locally indistinguishable and are connected to each other by braidings of the quasiparticles. This opens the path for non-Abelian braiding statistics.
{\em Second}, I  present a general conjecture on the properties of topological models arising from symmetrization of identical copies of a given topological model. I propose that the quasiparticles of the symmetrized model are constructed as symmetrizations of tensor products of the Abelian quasiparticles of the copied model.  The algebra characterizing the fusion and braiding rules is obtained through symmetrization of the tensor product of the algebras defining the copies.
{\em Third}, to illustrate the theory I analyze the model emerging from the symmetrization of two copies of the toric code model \cite{Kitaev2003}, a seminal example of an Abelian topological model. In order to characterize the arising symmetrized model, I use recent results on the description of topological models through the entanglement properties of the corresponding ground states 
\cite{KitaevPreskill,LevinWen,Vishwanath1,Vishwanath2,Vishwanath3}. Within this framework, the quasiparticles statistics and braiding can be derived by analyzing the basis of ground states with minimum entanglement entropy on a torus \cite{Vishwanath1,Vishwanath2,Vishwanath3}. Following this procedure, I analytically obtain the non-trivial fusion and braiding rules of the symmetrized model, proving its non-Abelian character and showing in this case the validity of the conjecture.

\section{SYMMETRIZED STATES AND NON-ABELIAN STATISTICS: INTUITION}

Let me consider a lattice system of local degrees of freedom  characterized by the Hilbert space $\mathcal{H}$. Let $\ket{\Phi}$ denote a collective state of such a many-body system. I further consider the state resulting from the symmetrization of $k$ identical copies of $\ket{\Phi}$ in the form:
\begin{equation}
\ket{\Psi}\propto \mathbf{P} \,
\left(
\ket{\Phi} \otimes \ket{\Phi} \ldots \otimes \ket{\Phi}
\right).
\label{Symmetrized1}
\end{equation}
Here, the projector $\mathbf{P}=\prod_\ell P_\ell$ is a product of local projectors that map the tensor product of the $k$ identical local degrees of freedom $\mathcal{H}^{\otimes k}$
onto a new degree of freedom $\mathcal{H}^S$, which is symmetric under exchange of any two of them:
\begin{eqnarray}
P: \mathcal{H}^{\otimes k}  \longrightarrow \mathcal{H}^S, \,\,\text{with} \,\,\,
P=\prod_{i<j}^{k}\frac{1+\mathsf{SWAP}_{ij}}{2},
\label{Projector}
\end{eqnarray}
and $\mathsf{SWAP}_{ij}$ being the permutation operator between copies $i$ and $j$.
The symmetrized state (\ref{Symmetrized1}) exhibits, by construction, a global hidden order associated to the internal organization of the local degrees of freedom in identical indistinguishable copies of the same many-body state. 

\begin{figure}[t]
\includegraphics[width=\linewidth]{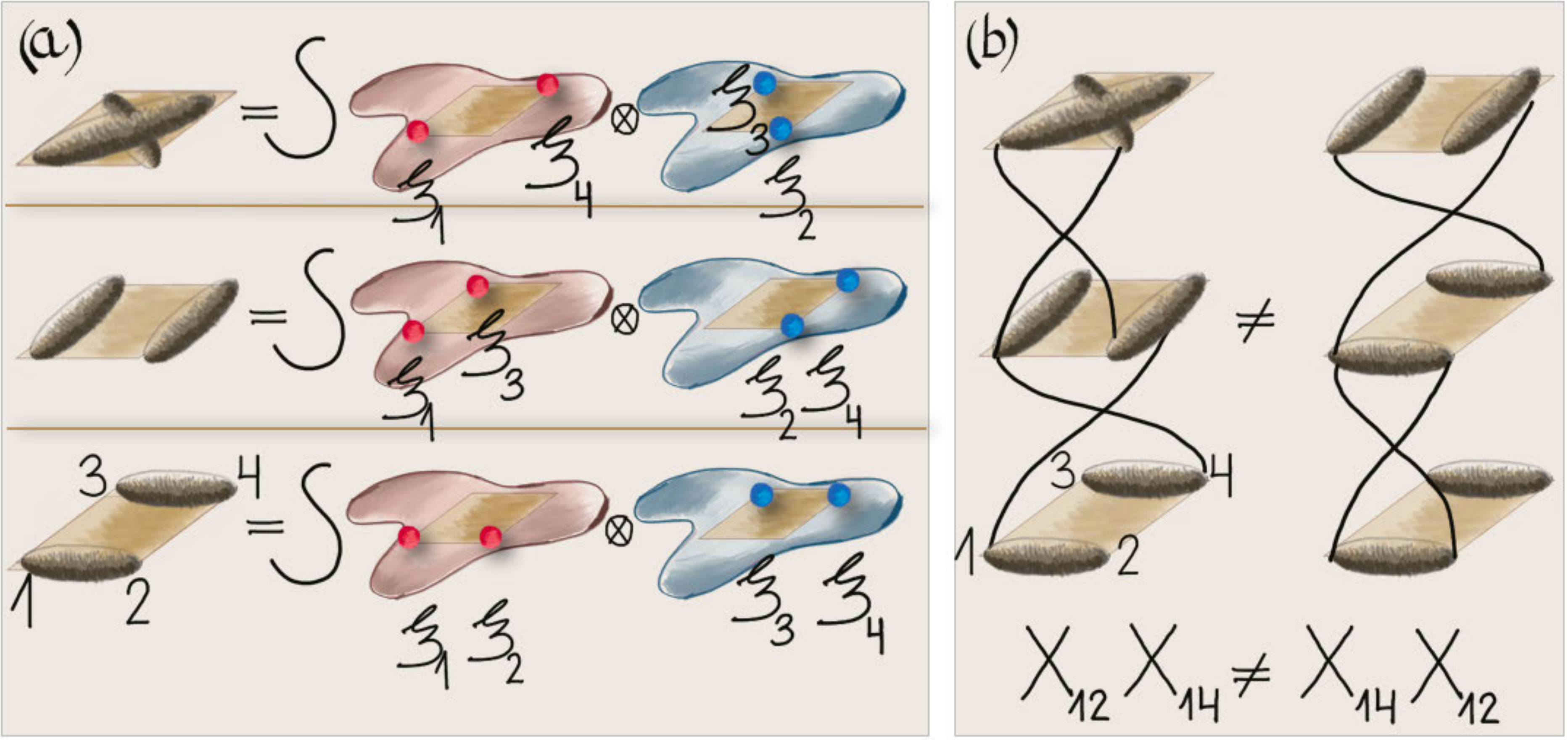}
\caption{\textbf{Topological degeneracy and non-Abelian braiding from symmetrization}. (\textbf{a}) A subspace of topologically distinct vacua's are compatible with the same positions of the quasiparticles. They correspond to different assignments of the quasiparticles to each of the identical copies. (\textbf{b}) Different states are connected to each other through braidings of the quasiparticles. The composition of braids is non-commutative.
}
\label{Braiding}
\end{figure}

Let me intuitively argue how quasiparticles with non-Abelian statistics can emerge from such a pattern of many-body entanglement. If we assume that each copy exhibits localized quasiparticles as elementary excitations, it is reasonable to expect that 
the elementary excitations of the symmetrized state are constructed by creating quasiparticles in each of the copies and symmetrizing. If the spatial positions of the quasiparticles are fixed, we still have freedom to assign them to each of the identical copies in different ways. For instance, for the case of two copies and four quasiparticles at positions $\xi_1, \xi_2, \xi_3, \xi_4$,
the following assignments are possible (Fig.\,\ref{Braiding}):
\begin{eqnarray}
&&\ket{\Psi_{(\xi_1\xi_2)(\xi_3\xi_4)}}\propto \mathbf{P} \,
(\ket{\Phi_{\xi_1\xi_2}}\otimes \ket{\Phi_{\xi_3\xi_4}})\nonumber\\
&&\ket{\Psi_{(\xi_1\xi_3)(\xi_2\xi_4)}}\propto \mathbf{P} \,
(\ket{\Phi_{\xi_1\xi_3}} \otimes \ket{\Phi_{\xi_2\xi_4}}) \nonumber\\
&&\ket{\Psi_{(\xi_1\xi_4)(\xi_2\xi_3)}}\propto \mathbf{P} \,
(\ket{\Phi_{\xi_1\xi_4}} \otimes \ket{\Phi_{\xi_2\xi_3}}).
\label{Excitations}
\end{eqnarray}
Here, the states $\ket{\Phi_{\xi_i\xi_j}}$ correspond to excited states of a copy with quasiparticles at positions $\xi_i$ and $\xi_j$.
The symmetrized states (\ref{Excitations}) are locally indistinguishable, since the positions of the quasiparticles are the same in all of them.
But they are globally distinct, for they correspond to different internal (hidden) pairings of the quasiparticles in each of the copies. Moreover, by braiding the quasiparticles the internal pairing can change, so that one state is transformed into another one (Fig.\,\ref{Braiding}). The composition of braids is non-commutative. If the operator that braids quasiparticles at positions $i$ and $j$ is denoted by $\mathsf{X}_{ij}$, we have, for example, that
\begin{eqnarray}
\mathsf{X}_{12}\mathsf{X}_{14}\ket{\Psi_{(\xi_1\xi_2)(\xi_3\xi_4)}}&=&\ket{\Psi_{(\xi_1\xi_4)(\xi_2\xi_3)}}\nonumber\\
\neq \mathsf{X}_{14}\mathsf{X}_{12}\ket{\Psi_{(\xi_1\xi_2)(\xi_3\xi_4)}}&=&\ket{\Psi_{(\xi_1\xi_3)(\xi_2\xi_4)}}.
\end{eqnarray}
We see how, due to symmetrization, a topological degeneracy seems plausible to arise in the subspace of quasiparticles, leading to a non-Abelian algebra of exchanges.

\section{Conjecture on symmetrized anyon models}
Topological models feature a ground state degeneracy that depends on the topology of the space on which they are defined \cite{Wen90,Kitaev2003}. Remarkably, the set of ground states on a torus already encodes complete information about the generalized statistics of the gapped quasiparticles
\cite{KitaevPreskill,LevinWen,Vishwanath1,Vishwanath2}. As proposed in \cite{Vishwanath1}, analyzing the entanglement properties of this set of ground states allows us to determine the braiding and fusion rules of the anyon model. One starts by noticing that the number of ground states on a torus corresponds to the number of distinct topological charges or anyonic quasiparticle types defining the model. Next, one identifies the states with {\em minimum entanglement entropy} for a bipartition of the torus along a given cycle. These states can be identified with the quasiparticles of the topological phase \cite{Vishwanath1}. Finally, the modular $S$-matrix, whose elements encode the mutual statistics of the quasiparticles, is obtained by relating bases of states with minimum entanglement entropy for different bipartitions of the torus associated to different cycles.

\subsection{Preliminaries: the Abelian model serving as a copy}

Let me consider an Abelian topological model with $d$ distinct topological charges labeled by 
\begin{equation}
\{q_1,\ldots,q_i,\ldots, q_d\}.
\label{LabelsCharges}
\end{equation}
Let this model describe a lattice of $N$ local degrees of freedom characterized by the local Hilbert space $\mathcal{H}$. For example, in the case of a lattice of spin-$\frac{1}{2}$ spins, this local Hilbert space is $\mathcal{H}\simeq\mathbbm{C}^2$. 

Let me consider the manifold of ground states of this model on a torus. For a bipartition of the torus along a given cycle,  I will consider the basis of states with minimum entanglement entropy for that bipartition:
\begin{equation}
\{\ket{\Phi_{q_1}},\ldots,\ket{\Phi_{q_i}},\ldots, \ket{\Phi_{q_d}}\}
\subset \mathcal{H}^{\otimes N}.
\label{MESAbelian}
\end{equation}
These states can be identified with the quasiparticles (\ref{LabelsCharges}) of the topological model.

The ground state subspace on the torus defines an effective Hilbert space of dimension $d$ that I will denote by  $\mathcal{H}_\mathcal{G}$.  It is important to distinguish between the local Hilbert space $\mathcal{H}$, which characterizes the local physical degrees of freedom of the topological model, and the {\em global} Hilbert space $\mathcal{H}_\mathcal{G}$, which describes the subspace of many-body ground states:
\begin{equation}
\mathcal{H}^{\otimes N} \supset \{\ket{\Phi_{q_i}}\} \simeq \mathcal{H}_\mathcal{G}.
\end{equation}
It is useful to define a basis of states in $\mathcal{H}_\mathcal{G}$ in one to one correspondence with the basis of mimimum entanglement entropy states in (\ref{MESAbelian}). I will denote these states by:
\begin{equation}
\{\ket{q_1},\ldots,\ket{q_i},\ldots, \ket{q_d}\}
\subset \mathcal{H}_\mathcal{G}.
\end{equation}

The fusion rules of the topological model can be derived from the algebra of operators connecting the states $\ket{q_i}$ to the vacuum state, which we will assume corresponds to the state  $\ket{q_1}$.  If we have that:
\begin{equation}
\ket{q_i}=X_i\ket{q_1}, \,\, \text{and}\,\,\,\, X_iX_j=\sum_k N^k_{ij}\,X_k,
\end{equation}
where each $N^k_{ij}$ is a nonnegative integer and the sum is over the complete
set of labels, it follows that the fusion rules are given by:
\begin{equation}
q_i \times q_j =\sum_k N^k_{ij}\,q_k.
\end{equation}
The modular $S$ matrix is derived by relating the bases of minimum entanglement entropy states for two different bipartitions of the torus.
If I denote by $\widetilde{\ket{\Phi_{q_i}}}$ the basis of states with minimum entanglement entropy for a bipartition along the other cycle of the torus, and by $\ket{\widetilde{q_i}}$ the corresponding basis of states in 
$\mathcal{H}_\mathcal{G}$, the elements of the modular $S$ matrix are given by:
\begin{equation}
S_{ij}=\langle \Phi_{q_i} \vert \widetilde{\Phi_{q_j}} \rangle=
\langle q_i \vert \widetilde{q_j} \rangle.
\end{equation}

Let me illustrate the definitions above with the case of a $\mathbbm{Z}_2$ Abelian model. In this case we have two distinct topological charges:
\begin{equation}
\{q_1\equiv 1,q_2\equiv e\}.
\end{equation}
Therefore the ground state manifold has dimension $d=2$ and $\mathcal{H}_\mathcal{G}$ is isomorphic to the Hilbert space of a spin-$\frac{1}{2}$ spin. The basis of ground states with minimum entanglement entropy for a given bipartition are in one to one correspondence with a certain basis in $\mathbbm{C}^2$:
\begin{eqnarray}
\mathcal{H}^{\otimes N} \ni \ket{\Phi_{1}} &\longleftrightarrow& \ket{1} \in \mathbbm{C}^2 \nonumber\\
\mathcal{H}^{\otimes N} \ni \ket{\Phi_{e}} &\longleftrightarrow& \ket{e} \in \mathbbm{C}^2.
\end{eqnarray}
The states of this basis are related to each other by an operator $X$, which squares to identity:
\begin{equation}
\ket{1} \xrightarrow{X}\ket{e}, \hspace{0.3cm}X^2=\mathbbm{1}.
\end{equation}
 The fusion rules directly follow from this property:
\begin{eqnarray}
\mathbbm{1} \cdot  \mathbbm{1}= \mathbbm{1} &\longrightarrow& 1 \times 1 =1 \nonumber\\
X \cdot \mathbbm{1} = X &\longrightarrow& e \times 1 =e \nonumber\\
X \cdot X = \mathbbm{1} &\longrightarrow& e \times e =1.
\end{eqnarray}

\subsection{Conjecture: Non-Abelian symmetrized model}

Let me  define a topological model characterized by a set of ground states on the torus that are obtained as symmetrization of two ground states of the previous Abelian model in the form:
\begin{eqnarray}
\ket{\Psi(q_i,q_j)}\equiv\mathbf{P}\left(\ket{\Phi_{q_i}}\otimes\ket{\Phi_{q_j}}\right).
\label{SymStates1}
\end{eqnarray}
By construction, the model describes a lattice of local degrees of freedom characterized by the Hilbert space 
$\mathcal{H}^S \subset \mathcal{H}\otimes\mathcal{H}$, which is symmetric under exchange of the local degrees of freedom $\mathcal{H}$.

Since by definition $\ket{\Psi(q_i,q_j)}=\ket{\Psi(q_j,q_i)}$, there will be $D=d(d+1)/2$ independent states of the form (\ref{SymStates1}). These states will not be in general orthogonal to each other. 
They define a ground state subspace of dimension $D$. The corresponding topological model has therefore $D$ topological charges that will be denoted by:
\begin{equation}
\{\mathcal{Q}_{ij}\}_{ i,j=1,\ldots, d, \,\,i<j}.
\end{equation}
I conjecture that this model is a non-Abelian topological model with the following properties:

i) {\em Global Hilbert space}. The global Hilbert space spanned by the many-body symmetrized states (\ref{SymStates1})
is isomorphic to 
$\mathcal{H}_\mathcal{G}^S \subset \mathcal{H}_\mathcal{G}\otimes \mathcal{H}_\mathcal{G}$, which is the subspace of the tensor product states in 
$\mathcal{H}_\mathcal{G}\otimes \mathcal{H}_\mathcal{G}$
that are symmetric under exchange of the two global degrees of freedom $\mathcal{H}_\mathcal{G}$:
\begin{eqnarray}
(\mathcal{H}^{S})^{\otimes N}\supset\{\ket{\Psi(q_i,q_j)}\} \simeq \mathcal{H}_\mathcal{G}^S \nonumber\\
\mathcal{H}_\mathcal{G}^S=P_\mathcal{G}(\mathcal{H}_\mathcal{G} \otimes \mathcal{H}_\mathcal{G}),
\end{eqnarray}
where $P_\mathcal{G}=\frac{1+\mathsf{SWAP}_\mathcal{G}}{2}$ and $\mathsf{SWAP}_\mathcal{G}$ exchanges the two global degrees of freedom $\mathcal{H}_\mathcal{G}$.

ii) {\em Topological charges}. The basis of ground states with minimum entanglement entropy, which correspond to the topological charges of the model, are obtained as linear combinations of the states (\ref{SymStates1}). I conjecture that this basis is in one to one correspondence with the basis in $\mathcal{H}_\mathcal{G}^S$ formed by the states:
\begin{eqnarray}
\ket{q_i \odot q_j}=\alpha_{ij}\left(\ket{q_i}\otimes\ket{q_j}+\ket{q_j}\otimes\ket{q_i}\right)\in 
\mathcal{H}_\mathcal{G}^S,
\,\,\,
\label{TopCharges}
\end{eqnarray}
where $\alpha_{ij}^{-1}=\sqrt{1+\delta_{ij}}$, with $\delta_{ij}$ being the Kronecker delta. Here, the states $\ket{q_i}$ define the basis of states in $\mathcal{H}_\mathcal{G}$ in one to one correspondence with the topological charges of the Abelian model.

iii) {\em Fusion rules}. The fusion rules are determined by the algebra of operators that create the 
charges $\ket{q_i \odot q_j}$ out of the vacuum $\ket{q_1 \odot q_1}$, with $\ket{q_1}$ denoting the vacuum of the Abelian model.

iv) {\em $S$-matrix}. If the Abelian model is characterized by a modular $S$-matrix with elements 
$S_{ij}$,
the $S$-matrix of the symmetrized model is given by:
\begin{eqnarray}
\mathcal{S}_{ij,i'j'}&=&\langle q_i \odot q_j   \vert \widetilde{q_{i'} \odot q_{j'}} \rangle\nonumber\\
&=&\frac{1}{2}(S_{ii'}S_{jj'}+S_{ij'}S_{ji'}).
\end{eqnarray}
\begin{figure}[t]
\includegraphics[width=\linewidth]{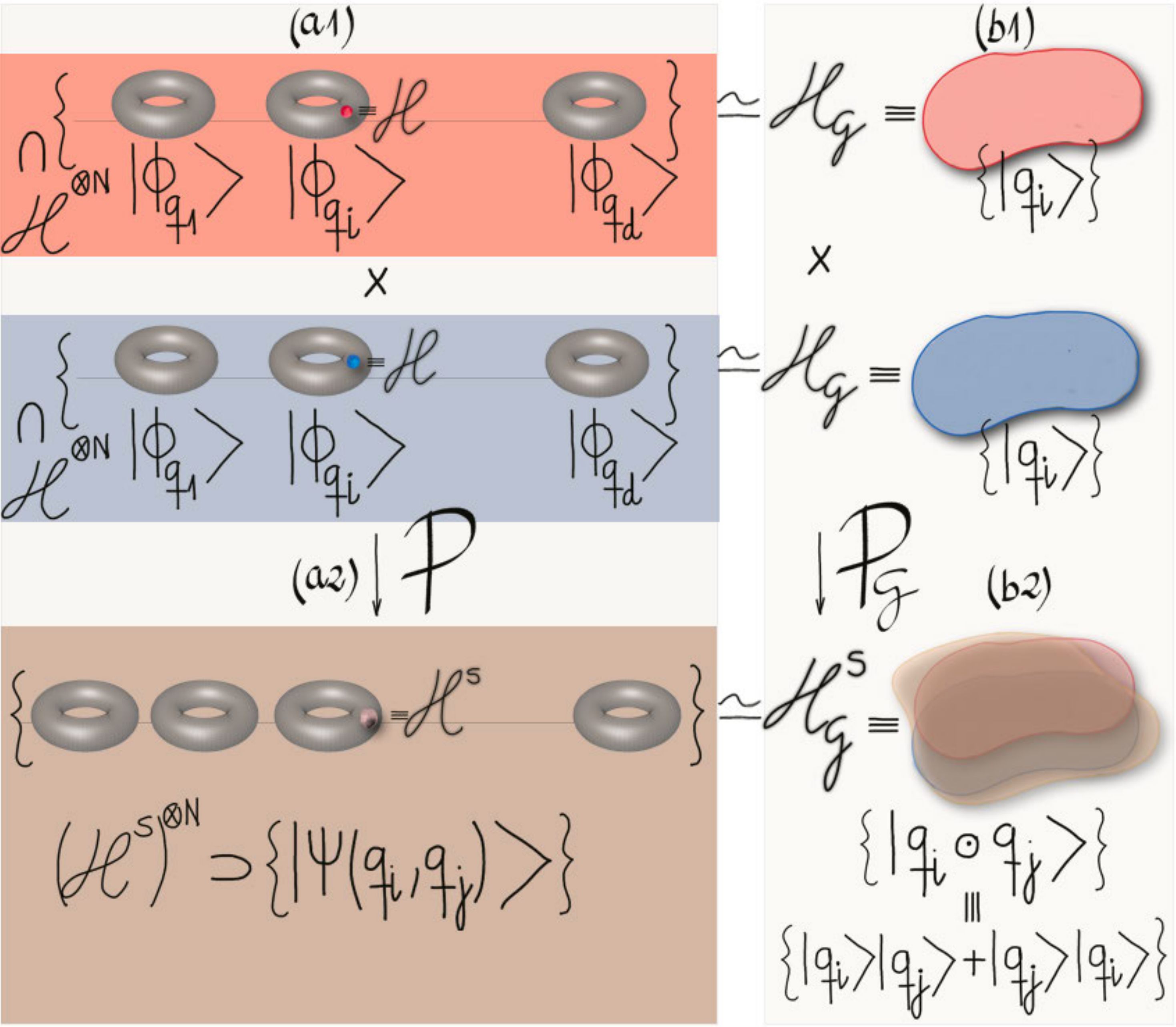}
\caption{\textbf{Local and global symmetrizations}. (\textbf{a}1) 
Manifold of ground states on the torus of an Abelian model describing a lattice of local degrees of freedom characterized by the Hilbert space $\mathcal{H}$. This manifold is isomorphic to the Hilbert space $\mathcal{H}_\mathcal{G}$ (\textbf{b}1). The many-body ground states $\ket{\Phi_{q_i}}$ are in one to one correspondence with states $\ket{q_i}$ in $\mathcal{H}_\mathcal{G}$. They define the topological charges of the Abelian model.
(\textbf{a}2) Two ground states of the Abelian model are symmetrized through the projector $\mathbf{P}$, which locally projects the tensor product $\mathcal{H} \otimes \mathcal{H}$ onto the local symmetric degree of freedom $\mathcal{H}^S$. The manifold of symmetrized states 
$\ket{\Psi(q_i,q_j)}$ on a torus defines the non-Abelian model. This manifold is isomorphic to $\mathcal{H}^S_\mathcal{G}$ 
(\textbf{b}2), which is symmetric under exchange of the two global degrees of freedom $\mathcal{H}_\mathcal{G}$. The topological charges of the symmetrized model, $\ket{q_i \odot q_j}$, are obtained as symmetrization of tensor products of Abelian charges $\ket{q_i}\otimes\ket{q_j}$. This symmetrization is carried out by the global projector $\mathbf{P}_\mathcal{G}$ (see text).
}
\label{GlobalLocalSym}
\end{figure}	
\subsection{Corollary: Symmetrization of two copies of an Abelian $\mathbbm{Z}_2$ model yields a non-Abelian Ising model}
From the conjecture above it follows that the model obtained by symmetrization of two identical copies of an Abelian $\mathbbm{Z}_2$ model is a non-Abelian Ising model \cite{Nayak2008,Preskill}.
First, we notice that the symmetrized model has $D=3$ topological charges:
\begin{eqnarray}
\{\mathcal{Q}_{11}\equiv \mathcal{I},\mathcal{Q}_{12}\equiv \sigma,\mathcal{Q}_{22}\equiv \psi\}.
\end{eqnarray}
According to point i) of the conjecture, the space of ground states is isomorphic to the symmetric subspace of two 
spin-$\frac{1}{2}$ spins, which corresponds to the Hilbert space of a spin-$1$ spin:
\begin{eqnarray}
\{\ket{\Psi_{11}},\ket{\Psi_{1e}},\ket{\Psi_{ee}}\}\simeq \mathcal{H}_\mathcal{G}^S \simeq \text{spin-1}.
\end{eqnarray}
Next, point ii) of the conjecture states that the basis of states in $\mathcal{H}_\mathcal{G}^S$ in correspondence with the topological charges of the symmetrized model are obtained as symmetrizations of products of states in $\mathcal{H}_\mathcal{G}$ corresponding to the topological charges of the Abelian model:
\begin{eqnarray}
\mathcal{I}\leftrightarrow\ket{1\odot 1}&=&\ket{1}\otimes\ket{1}\nonumber\\
\sigma\leftrightarrow\ket{1\odot e}&=&\frac{1}{\sqrt{2}}(\ket{1}\otimes\ket{e}+\ket{1}\otimes\ket{e})\nonumber\\
\psi\leftrightarrow\ket{e\odot e}&=&\ket{e}\otimes\ket{e}.
\end{eqnarray}
Finally, the algebra of the spin-1 operators connecting the states above determines the fusion rules of the symmetrized model.
We have that:
\begin{eqnarray}
\ket{1\odot 1}&\xrightarrow{\sqrt{2}S_x}\ket{1\odot e}\longrightarrow& \ket{e\odot e}\nonumber\\
\ket{1\odot 1}&\xrightarrow {\,\,\mathcal{X}=2S_x^2-1\,\,}&\ket{e\odot e},
\end{eqnarray}
with $S_x\equiv (X\otimes \mathbbm{1} + \mathbbm{1} \otimes X)/2$ being an effective spin-1 operator component. Therefore we conclude that:
\begin{eqnarray}
\sqrt{2}S_x\sqrt{2}S_x=\mathbbm{1}+\mathcal{X} &\,\,\,\longrightarrow \,\,\, &\sigma \times \sigma=\mathcal{I}+\psi \nonumber\\
\sqrt{2}S_x\,\mathcal{X}=\sqrt{2}S_x &\,\,\,\longrightarrow \,\,\, &\sigma \times \psi=\sigma \nonumber\\
\mathcal{X}\,\mathcal{X}=\mathbbm{1}&\,\,\,\longrightarrow \,\,\,&\psi\times\psi=\mathcal{I}.
\end{eqnarray}
These fusion rules are non-Abelian, since the charge $\sigma$ has two fusion channels: $\mathcal{I}$ and $\psi$. They correspond to an Ising model \cite{Nayak2008,Preskill}.

Furthermore, it follows from the conjecture that the symmetrization of $k$ identical copies of a $\mathbbm{Z}_2$ Abelian model yields a $SU(2)_k$ \cite{Nayak2008,Preskill} non-Abelian topological model. Similarly to the discussion above, the effective global Hilbert space of such symmetrized model corresponds to the one of a spin-$\frac{k}{2}$ spin. The algebra of spin-$\frac{k}{2}$ operators yields non-Abelian fusion rules corresponding to an $SU(2)_k$ topological model.

\subsection{The core of the conjecture}
The conjecture above states that the symmetrization of the product of identical local degrees of freedom corresponding to two copies of a topological model implies a symmetrization of the two identical Hilbert spaces describing the many-body ground state subspaces. This correspondence between the local and global symmetrizations is the core of the conjecture. Once we have established such correspondence, the non-Abelian character of the symmetrized model arises in a natural way. As we have seen above for the case of two copies of a $\mathbbm{Z}_2$ model, the operators connecting states in the symmetrized global Hilbert space define an algebra that yields non-Abelian fusion rules. These non-Abelian properties arise from the restriction imposed by the symmetrization on the tensor product of operators defining the Abelian algebra. The crucial point of the conjecture is to state that a symmetrized algebra (which is naturally non-Abelian) has a physical realization: it corresponds to a subspace of many-body ground states obtained from two Abelian ground states through symmetrization of the local degrees of freedom.


Points i) and ii) constitute the core of the conjecture. To prove i) one needs to show the existence of a correspondece between the local symmetrization carried out by the projector $\mathbf{P}$ on the tensor product of local degrees of freedom $\mathcal{H}\otimes \mathcal{H}$, and the global symmetrization that projects the tensor product of global spaces 
$\mathcal{H}_\mathcal{G}\otimes\mathcal{H}_\mathcal{G}$ onto the symmetric global subspace 
$\mathcal{H}_\mathcal{G}^S$ (see Fig.\,\ref{GlobalLocalSym}).  Point ii) further requires to demonstrate that 
the quasiparticles of the symmetrized model are obtained from symmetrization in the global subspace of products of charges of the Abelian model that serves as a copy. This can be done using the correspondence i). Points iii) and iv) follow directly from i) and ii).

Here, I will consider the case of two copies of the toric code model and prove that the emerging symmetrized model satisfies the conjecture above. First, I will establish an isomorphism between the manifold of many-body symmetrized states and the symmetric global Hilbert space. Then I will show that the topological charges of the symmetrized model are given by (\ref{TopCharges}). Finally, I will obtain the fusion rules and the $S$-matrix of the model, demonstrating its non-Abelian character. 

\section{CHARGES, FUSION RULES AND S-MATRIX OF THE TORIC CODE MODEL}
The toric code model \cite{Kitaev2003} is a seminal example of an Abelian topological model. 
It describes 
spin-$\frac{1}{2}$ spins sitting at the edges of a square two-dimensional lattice.  
The ground state (see Fig.\,\ref{ToricCode}) is a superposition with equal weight of all possible spin states in which up-spins are arranged along a closed loop configuration $\mathcal{L}$:
\begin{equation}
\ket{\Phi} \propto 
\sum_{\{\mathcal{L}\}} 
\ket{\mathcal{L}}. 
\label{KitaevState}
\end{equation}
Here,
$\ket{\mathcal{L}} = \prod_{\ell \in \mathcal{L}} 
\sigma_{\ell}^x 
\ket{\textrm{vac}}$, 
and
$\ket{\textrm{vac}} =\bigotimes_{\ell=1}^N\ket{\downarrow}_\ell$.
\subsection{Ground state subspace on the torus}

\begin{figure}[t!]
\includegraphics[width=\linewidth]{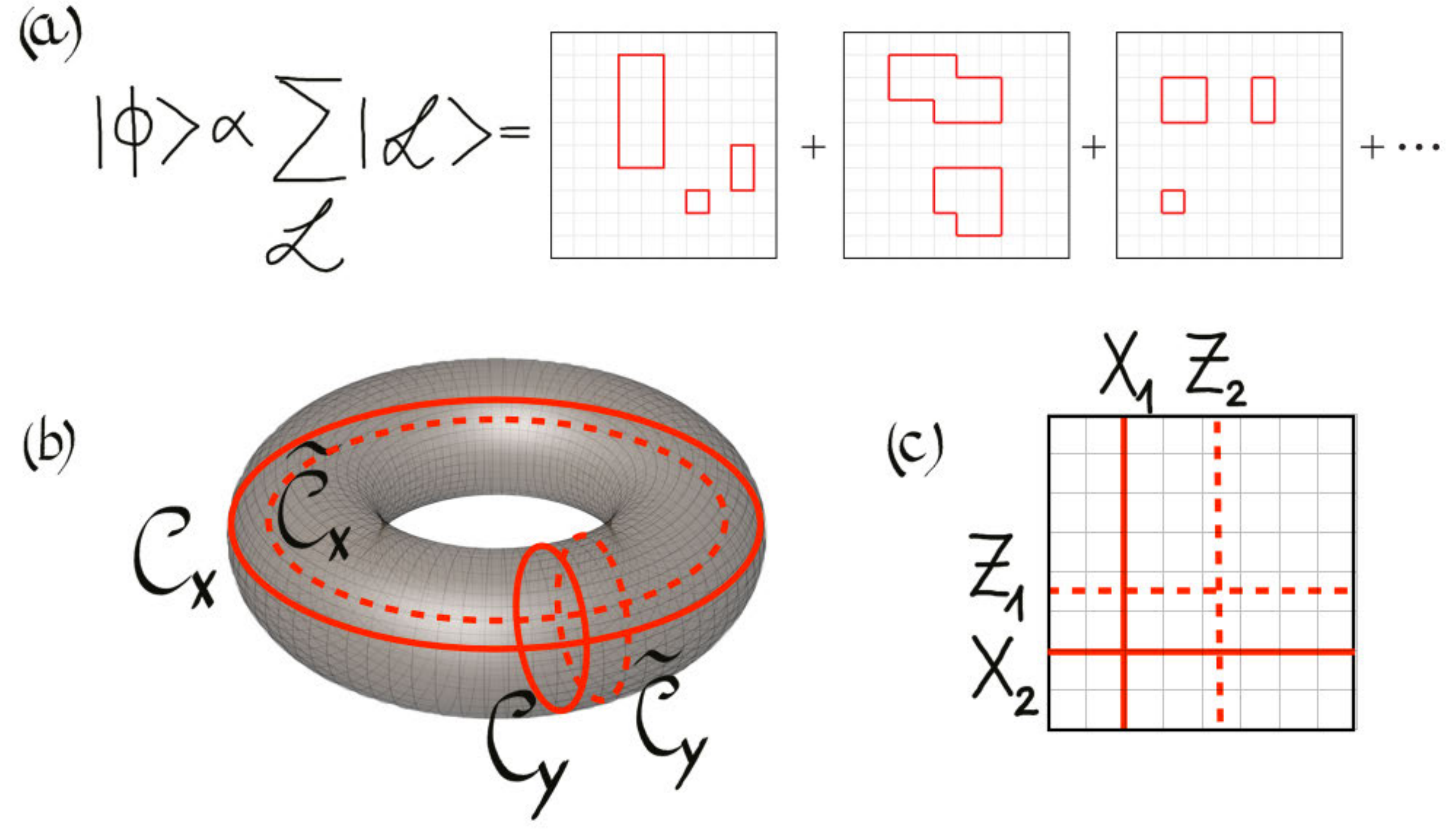}
\caption{\textbf{Toric code model}. The toric code model describes $\frac{1}{2}$-spin spins sitting at the edges of a square lattice. (\textbf{a}) The ground state is a superposition of closed loops configurations, where the presence or absence of a line segment at a given edge represents, respectively, the state up or down of the spin-$\frac{1}{2}$ spin. (\textbf{b}) On the torus there are four degenerate ground states, corresponding to the two classes of non-contractible loops $\mathcal{C}_x,\mathcal{C}_y$ ($\widetilde{\mathcal{C}}_x,\widetilde{\mathcal{C}}_y$) in the normal 
(dual) lattice. (\textbf{c}) The string operators $X_1, Z_1, X_2, Z_2$ (see text) define an effective 
two-qubit algebra in the space of ground states on the torus. 
}
\label{ToricCode}
\end{figure}

The properties of the topological model (topological charges, fusion and braiding rules) are encoded in the manifold of ground states on a torus. There are $4$ ground states:
\begin{eqnarray}
\ket{\Phi},\,\,X_1\ket{\Phi}, 
\,\,
X_2\ket{\Phi},
\,\,X_1X_2\ket{\Phi},
\label{4states}
\end{eqnarray}
where 
$X_{1(2)}=\prod_{\ell \in \mathcal{C}_{x(y)}} \sigma_{\ell}^x$, and $\mathcal{C}_{x(y)}$ are the two different non-contractible loops (see Fig.\,\ref{ToricCode}). 
These states define an effective Hilbert space $\mathcal{H}_\mathcal{G}$ of dimension $d=4$. It is straightforward to see that this Hilbert space corresponds to the one of two effective spin-$\frac{1}{2}$ spins (the one of two qubits):
\begin{equation}
\mathcal{H}_\mathcal{G} 
\simeq \mathbbm{C}^2 \otimes \mathbbm{C}^2 \simeq \frac{1}{2}\otimes\frac{1}{2}.
\end{equation}
If we define dual operators 
$Z_{1(2)}=\prod_{\ell \in \widetilde{\mathcal{C}}_{x(y)}} \sigma_{\ell}^z$,
with $\widetilde{\mathcal{C}}_{x(y)}$ being non-contractible loops in the dual lattice, we have:
\begin{equation}
\{X_{1(2)},Z_{1(2)}\}=0.
\label{Vacuum}
\end{equation}
The operators 
$Z_{1(2)}(X_{1(2)})$ therefore represent effective $z(x)$-Pauli matrices. The many-body states in 
(\ref{4states}) are in one to one correspondence with states of the two spin-$\frac{1}{2}$ spins for which both spins have well defined $z$-component:
\begin{eqnarray}
\begin{array}{crccc}
&\ket{\Phi}& \longleftrightarrow & \ket{+\hat{z}}\ket{+\hat{z}}\equiv\ket{1}&\\
&X_2\ket{\Phi} &  \longleftrightarrow & \ket{+\hat{z}}\ket{-\hat{z}}\equiv\ket{2}&\\
&X_1\ket{\Phi} & \longleftrightarrow & \ket{-\hat{z}}\ket{+\hat{z}}\equiv\ket{3}&\\
\mathcal{H}^{\otimes N}\ni& X_1X_2\ket{\Phi}&\longleftrightarrow & \ket{-\hat{z}}\ket{-\hat{z}}
\equiv\ket{4} & \in \mathcal{H}_\mathcal{G}.
\label{zzBasis}
\end{array}
\end{eqnarray}
Here, $\ket{\pm\hat{z}}\in \mathbbm{C}^2$ and $\sigma_z\ket{\pm\hat{z}}=\pm\ket{\pm\hat{z}}$
\footnote{\label{Note1}Along the manuscript I abuse of notation and use $Z$ and $X$ to  denote, respectively, the Pauli operators $\sigma_z$ and $\sigma_x$ acting on the effective qubits. Though this is not formally correct, since $Z$ and $X$ have been defined above as string operators acting on the many body system $\mathcal{H}^{\otimes N}$, this choice simplifies the notation when considering the symmetrized model.}.
\begin{figure}[t]
\includegraphics[width=\linewidth]{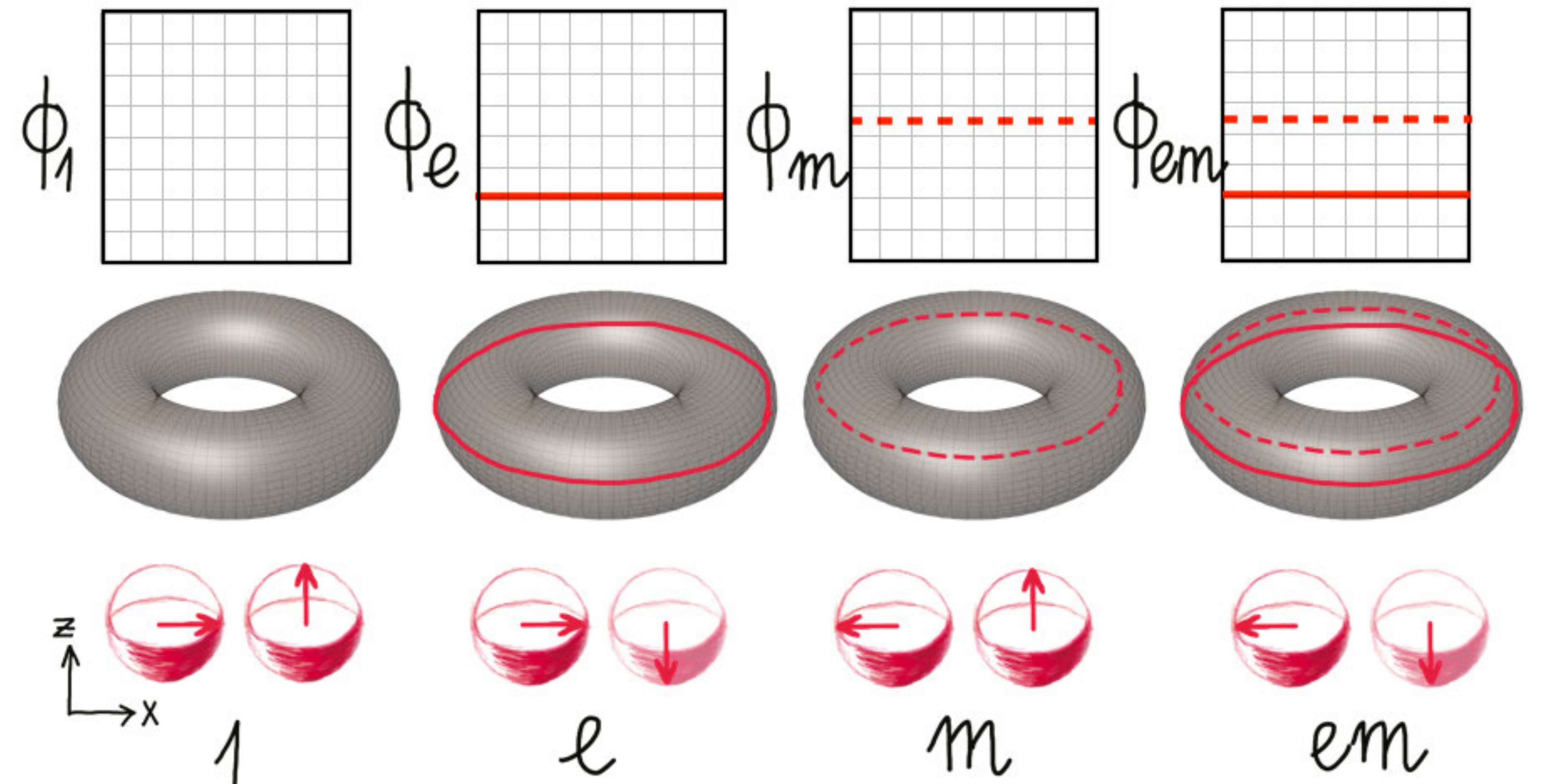}
\caption{\textbf{Topological charges of the toric code model}. The charges along the $x$-direction correspond to states of two qubits for which the first and the second qubit have well defined $x$-component and $z$-component, respectively. The qubit states are represented by vectors in the Bloch sphere. Starting from the vacuum state $\ket{1}$, an electric charge $\ket{e}$ is created by flipping the second qubit, whereas flipping the first qubit creates a magnetic charge $\ket{m}$.
}
\label{ChargesToricCode}
\end{figure}
\subsection{Topological charges and minimum entanglement entropy states}
The model has $d=4$ topological charges:
\begin{equation}
\{q_1\equiv 1,q_2\equiv e, q_3\equiv m, q_4\equiv em\},
\end{equation}
which correspond, respectively, to the vacuum, the electric charge, the magnetic charge and the composite of electric and magnetic charges.
The basis of ground states with minimum entanglement entropy, which correspond to the topological charges of the model, are obtained as linear combinations of the states (\ref{4states}). For a bipartition of the torus along the $x$-axis these states are \cite{Vishwanath1,Vishwanath2}:
\begin{eqnarray}
&\ket{\Phi_1},&
\ket{\Phi_e}\equiv X_2 \ket{\Phi_1}, \nonumber\\
&\ket{\Phi_m}\equiv Z_1\ket{\Phi_1}, & \ket{\Phi_{em}}\equiv X_2Z_1\ket{\Phi_{1}},
\label{4charges}
\end{eqnarray}
where $\ket{\Phi_1}\equiv\frac{1}{\sqrt{2}}(1+X_1)\ket{\Phi}$. 
They correspond to states of the two effective qubits for which the first qubit has well defined $x$-component, whereas the second has well defined $z$-component 
(see Fig.\,\ref{ChargesToricCode}):
\begin{eqnarray}
\begin{array}{clccc}
&\ket{\Phi_{1}}& \longleftrightarrow & \ket{+\hat{x}}\ket{+\hat{z}}\equiv\ket{q_1}&\\
&\ket{\Phi_{e}} &  \longleftrightarrow & \ket{+\hat{x}}\ket{-\hat{z}}\equiv\ket{q_2}&\\
&\ket{\Phi_{m}} & \longleftrightarrow & \ket{-\hat{x}}\ket{+\hat{z}}\equiv\ket{q_3}&\\
\mathcal{H}^{\otimes N}\ni& \ket{\Phi_{em}}&\longleftrightarrow & \ket{-\hat{x}}\ket{-\hat{z}}
\equiv\ket{q_4} & \in \mathcal{H}_\mathcal{G}.
\label{xzBasis}
\end{array}
\end{eqnarray}
Here, $\ket{\pm\hat{x}}\in \mathbbm{C}^2$ and $\sigma_x\ket{\pm\hat{x}}=\pm\ket{\pm\hat{x}}$.

I have introduced three different bases of states, which will be important when discussing the topological symmetrized model below. The basis of many-body states with minimum entanglement entropy, $\ket{\Phi_{q_i}} \in \mathcal{H}^{\otimes N}$, defined in (\ref{4charges}), is in one to one correspondence with the basis of states $\ket{q_i}$ in $\mathcal{H}_\mathcal{G}$ defined in (\ref{xzBasis}). Additionally, we have the basis of states $\ket{i}$ in $\mathcal{H}_\mathcal{G}$, defined in (\ref{zzBasis}), which is related to the basis of states $\ket{q_i}$ by applying the Hadamard transformation $\textsf{H}$ to the first qubit:
\begin{eqnarray}
\mathcal{H}^{\otimes N} \ni\ket{\Phi_{q_i}} \longleftrightarrow \ket{q_i}=\textsf{H}\otimes1\ket{i} \in \mathcal{H}_\mathcal{G}.
\label{GlobalCharges}
\end{eqnarray}

\subsection{Fusion rules and S-matrix}
The fusion rules can be obtained from (\ref{4charges}) taking into account the properties of 
the operators $Z_1$ and $X_2$:
\begin{eqnarray}
\begin{array}{lll}
e \times e =1 & m\times m=1  & e\times m=em \nonumber\\ 
em \times e= m & em \times m= e & em \times em=1.
\end{array}
\label{FusionRulesOne}
\end{eqnarray}
These fusion rules are trivial, with only one channel, indicating that the model is Abelian.

The modular $S$-matrix is derived by relating the bases of minimum entanglement entropy states for two different bipartitions of the torus. For a bipartition of the torus along the $y$-axis the basis of minimum entanglement entropy states is in one to one correspondence with a basis of states $\ket{\widetilde{q_i}}$, which is obtained from $\ket{q_i}$ by exchanging the two qubits:
\begin{eqnarray}
\ket{\widetilde{q_i}}=\mathsf{SWAP}\ket{q_i}.
\end{eqnarray}
Here, $\mathsf{SWAP}$ is the unitary operation exchanging the two qubits.
Therefore, we have:
\begin{eqnarray}
S_{ij}&=&\langle q_i \,\vert \,\widetilde{q}_j\rangle= 
\bra{q_i} \mathsf{SWAP} \ket{q_j}\nonumber\\
&=&\bra{i}\mathsf{H}\otimes1 \cdot\mathsf{SWAP}\cdot \mathsf{H}\otimes 1\ket{j}\nonumber\\
&=&\bra{i}\mathsf{SWAP}\cdot \mathsf{H}\otimes \mathsf{H}\ket{j}.
\end{eqnarray}


\section{SYMMETRIZED ANYON MODEL FROM TWO COPIES OF THE TORIC CODE}
\subsection{Spin-1 symmetrized state}
Let me consider the many-body state describing a lattice of spin-1 particles that is constructed by symmetrization of two identical copies of the toric code ground state in Eq.\,(\ref{KitaevState}):
\begin{eqnarray}
\ket{\Psi}=\mathbf{P}\left(\ket{\Phi}\otimes\ket{\Phi}\right).
\label{SymState}
\end{eqnarray}
Here, the projection $\mathbf{P}=\prod_\ell P_\ell$ is a product of local projectors that map the tensor product of two spins-$\frac{1}{2}$ spins onto the symmetric subspace of total spin-$1$:
\begin{eqnarray}
P_\ell:\mathcal{H}\otimes\mathcal{H}\simeq\frac{1}{2}\otimes\frac{1}{2}
\longrightarrow \mathcal{H}^{S}\simeq \text{spin-1},
\label{LocalProjector}
\end{eqnarray}
\begin{figure}[t]
\includegraphics[width=\linewidth]{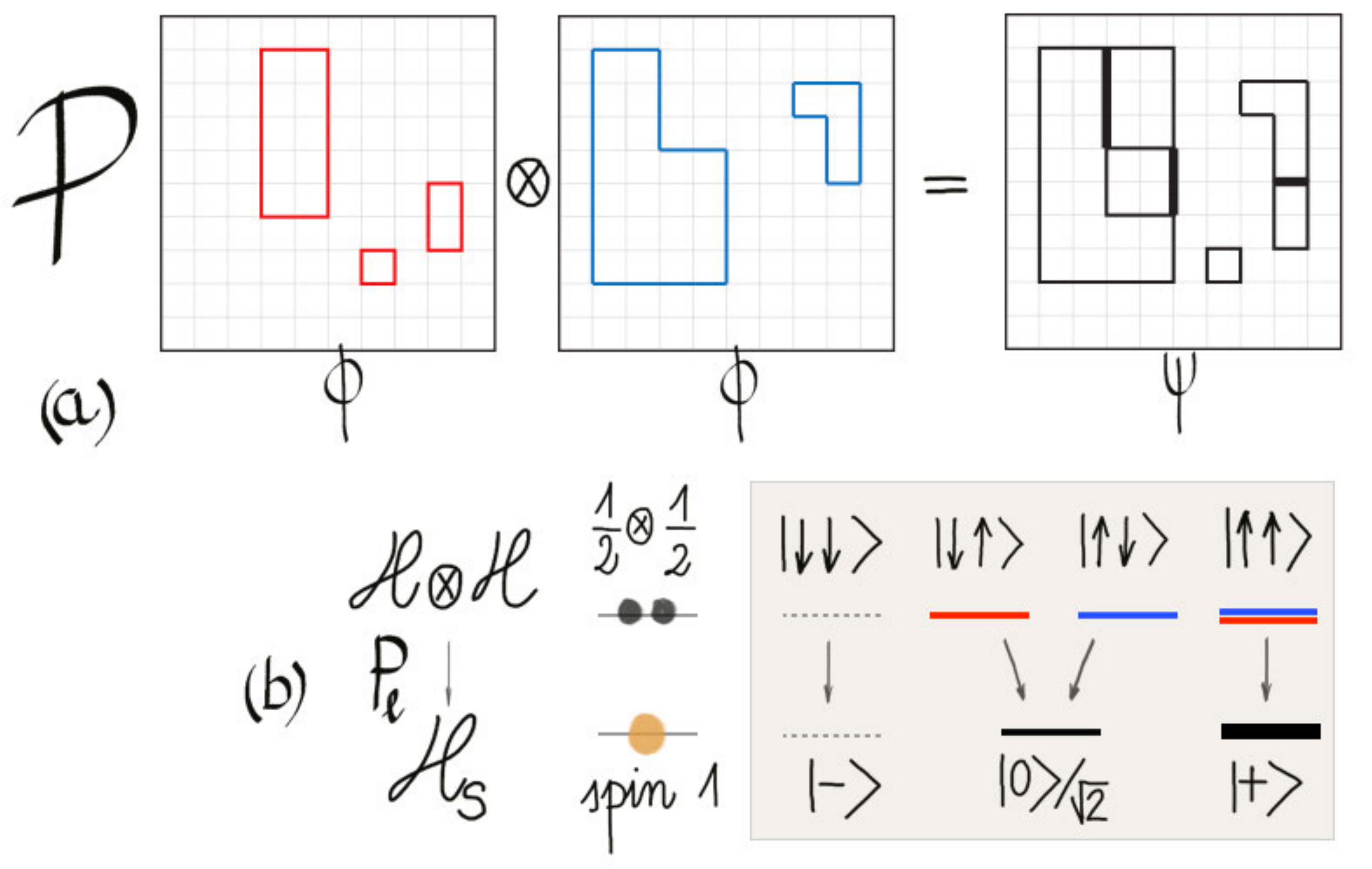}
\caption{\textbf{Spin-1 symmetrized state}. (\textbf{a}) Two copies of a toric code ground state $\Phi$ are merged by the projector $\mathbf{P}$, leading to the spin-1 state $\Psi$. (\textbf{b}) Local projection of two spin-$\frac{1}{2}$ spins onto a spin-$1$ spin. Spin-$\frac{1}{2}$ and spin-1 states are represented by line segments.
}
\label{Spin1State}
\end{figure}
In terms of local spin-$1$ operators, $S^\alpha_\ell$, $\alpha=x,y,z$, the state (\ref{SymState}) has the form:
\begin{eqnarray}
\ket{\Psi}\propto
\prod_p \left(1+B_{1p}\right) \sum_{\{\mathcal{L}\}} \ket{\mathcal{L}_1}.
\label{SymLoop}
\end{eqnarray}
Here, $\ket{\mathcal{L}_1}=\prod_{\ell \in \mathcal{L}} S^x_\ell \ket{-}$, 
with $\ket{-}=\bigotimes_{\ell=1}^N\ket{-}_\ell$, and $\ket{-}_\ell$ denotes the  state with minimum $z$-component of the $\ell$th spin-$1$ spin.
The plaquette operator 
\begin{equation}
B_{1p}=\prod_{\ell \in p} (2[S_{\ell}^x]^2-1)
\end{equation} 
acts on the four spins within a lattice plaquette $p$.
Using a language of strings, the three orthogonal states of the local spin-$1$ degree of freedom:
\begin{eqnarray} 
\ket{-}_\ell &&\nonumber\\
\frac{1}{\sqrt{2}}\ket{0}_\ell&=&S^x_\ell\ket{-}_\ell,\nonumber \\
\ket{+}_\ell&=& (2[S_{\ell}^x]^2-1)\ket{-}_\ell,
\end{eqnarray}
are mapped, respectively, onto no-segment, single-line segment, and double-line segment in the corresponding loop configuration (see Fig.\,\ref{Spin1State}). 
Within this language, the states $\ket{\mathcal{L}_1}$ are closed loop configurations of single lines, and the symmetrized state (\ref{SymLoop}) is a superposition of closed loops made of single lines, modulo plaquette moves $B_{1p}$ that exchange no-lines with double-lines.

For all the discussion that follows, the actual form of the symmetrized state in terms of spin-1 operators is not relevant. The properties of the symmetrized anyon model will be derived only taking into account the form (\ref{SymState}), using both the properties of the projector and the properties of the Abelian anyon model defining the copies.
\subsection{Symmetrized ground state subspace on the torus}
Let me consider the topological model that results from symmetrization of two identical toric code models.  
This model is defined through a set of ground states on the torus that are obtained as symmetrization of two ground states of the toric code model in the form:
\begin{eqnarray}
\ket{\Psi(q_i,q_j)}\equiv\mathbf{P}\left(\ket{\Phi_{q_i}}\otimes\ket{\Phi_{q_j}}\right),
\label{Symm States qq}
\end{eqnarray}
where the states $\ket{\Phi_{q_i}}$ are defined in (\ref{4charges}).
In the following I analyze the manifold of ground states (\ref{Symm States qq}) and characterize the corresponding symmetrized topological model.

\begin{figure}[t]
\includegraphics[width=\linewidth]{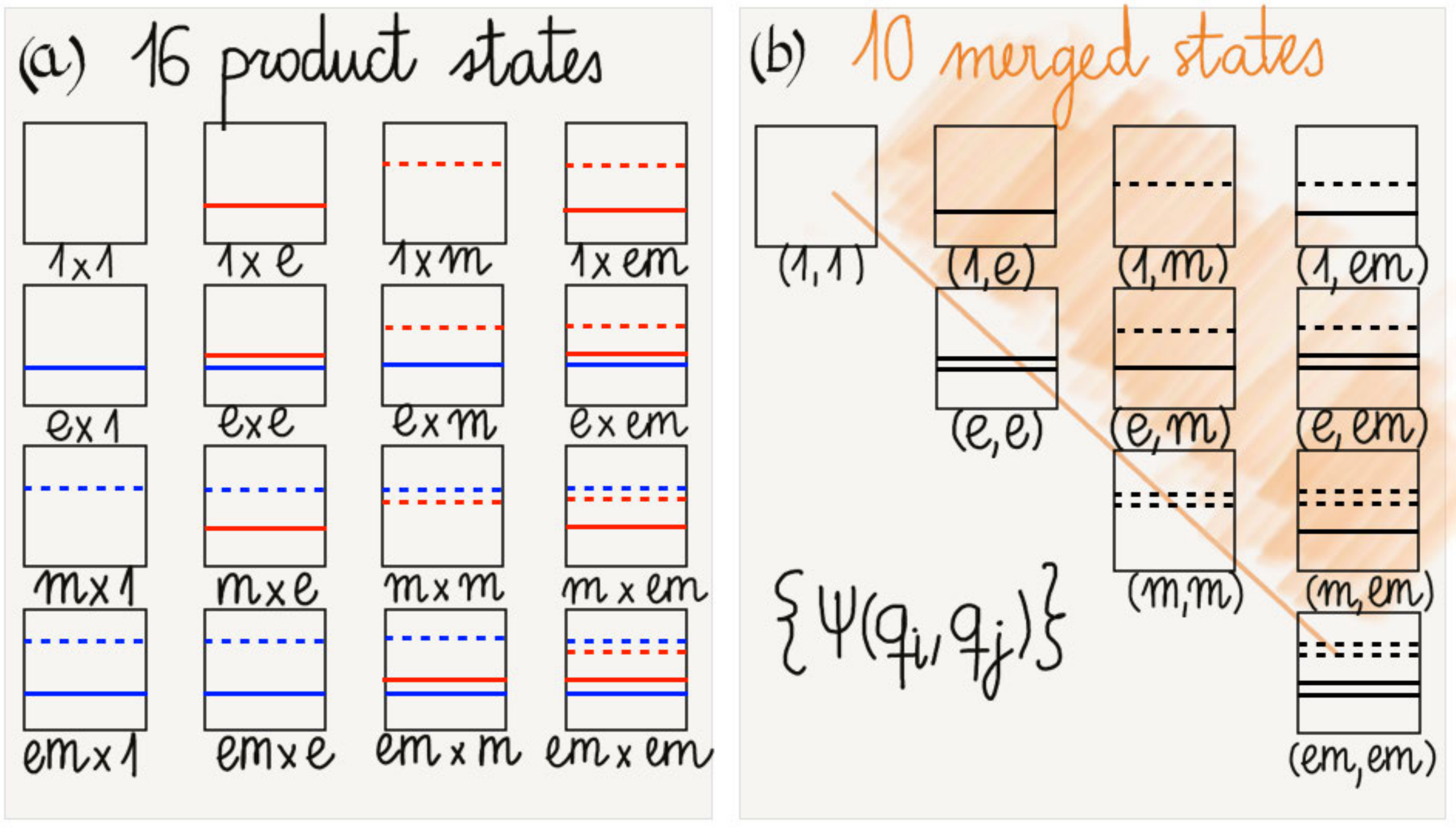}
\caption{\textbf{Symmetrized states on the torus}. The 16 product states 
$\Phi_{q_i} \otimes \Phi_{q_j}$ (\textbf{a}) corresponding to two copies of the toric code ground state subspace are mapped, after symmetrization through the projector 
$\mathbf{P}$, onto 10 linearly independent symmetrized states $\Psi(q_i,q_j)$ (\textbf{b}).
}
\label{SymStatesTorus}
\end{figure}

\begin{figure}[t]
\includegraphics[width=\linewidth]{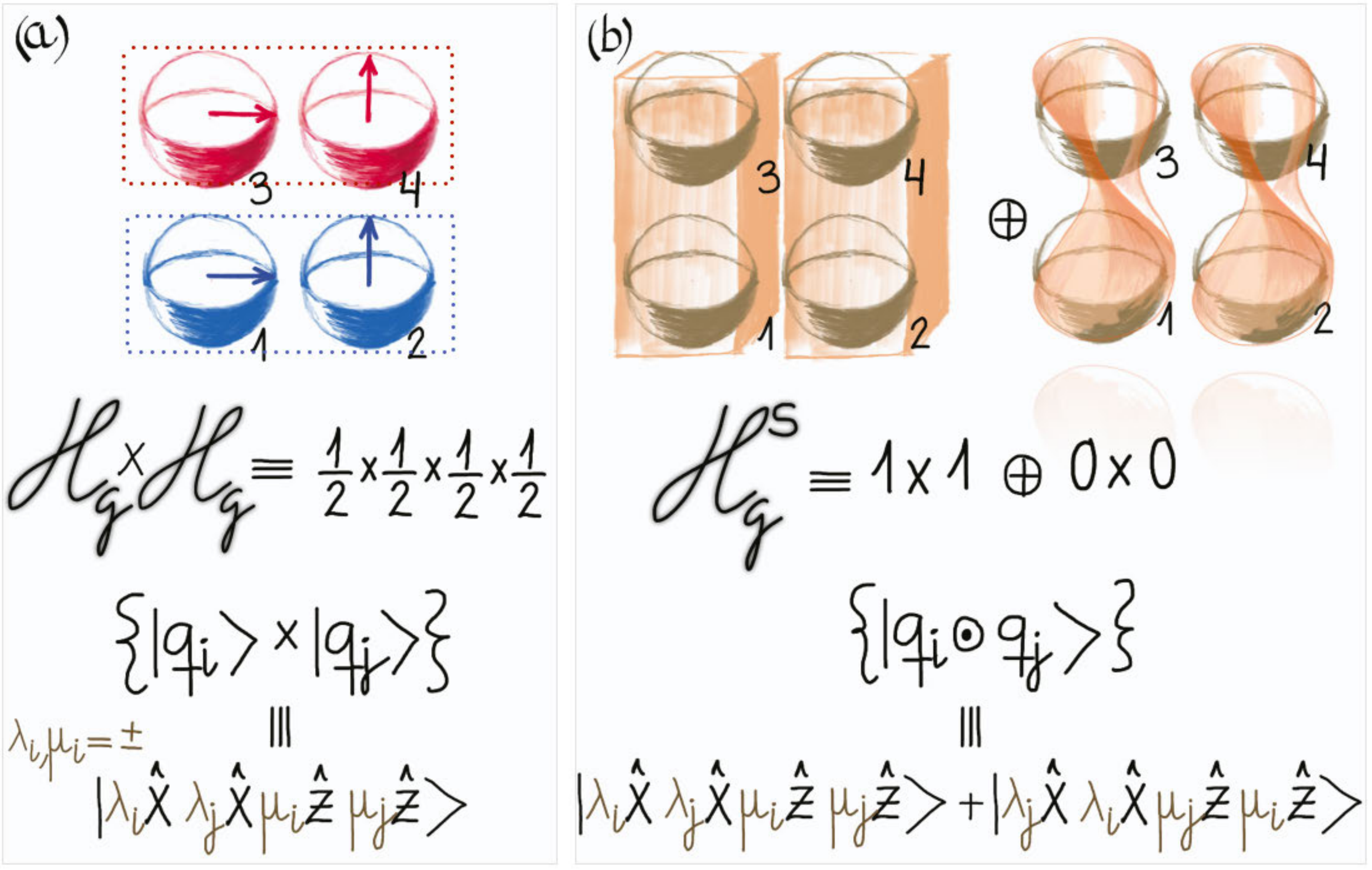}
\caption{\textbf{Four qubits symmetric subspace}. 
The tensor product  space of four qubits (\textbf{a}),
corresponding to two copies of the ground state manifold of the toric code model, is mapped after symmetrization onto a subspace (\textbf{b}) that is symmetric under exchange of the two pairs of qubits $\{1,2\}$ and $\{3,4\}$. The basis of product charges $\ket{q_i}\otimes \ket{q_j}$ is mapped onto the basis of symmetric charges $\ket{q_i\odot q_j}$.
}
\label{4Qubits}
\end{figure}

Before symmetrization, there are $16$ independent states of the form 
$\ket{\Phi_{q_i}} \otimes \ket{\Phi_{q_j}}$, corresponding to all possible tensor products of two toric code ground states (Fig.\,\ref{SymStatesTorus}a). The Hilbert space of these product states is isomorphic to the one of four qubits (Fig.\,\ref{4Qubits}a), with the correspondence:
\begin{eqnarray}
(\mathcal{H}\otimes\mathcal{H})^{\otimes N} \ni
\ket{\Phi_{q_i}}\otimes\ket{\Phi_{q_j}} 
\longleftrightarrow \ket{q_i} \otimes \ket{q_j} \in 
\mathcal{H}_\mathcal{G}\otimes\mathcal{H}_\mathcal{G} \nonumber.
\end{eqnarray}
If the qubits of the first (second) copy are denoted by $\{1,2\}$ ($\{3,4\}$), the states 
$\ket{q_i} \otimes \ket{q_j}$ correspond to the basis in which qubits   
$\{1,3\}$ have well defined $x$-component, whereas $\{2,4\}$ have well defined $z$-component:
\begin{eqnarray}
\ket{q_i}\otimes\ket{q_j}&=&
\ket{\lambda_i\hat{x}}_1\ket{\lambda_j\hat{x}}_3\ket{\mu_i\hat{z}}_2\ket{\mu_j\hat{z}}_4,
\label{ProductStates}
\end{eqnarray}
with $\lambda_{i(j)},\mu_{i(j)}=\pm$.

The symmetrization carried out by the operator $\mathbf{P}$ makes some of the tensor product states indistinguishable, leaving $D=10$ linearly independent symmetrized states  
of the form (\ref{Symm States qq}) (Fig.\,\ref{SymStatesTorus}b). I will show below that 
the subspace spanned by these states  is isomorphic to the subspace of the Hilbert space of four qubits that is symmetric under exchange of the two pairs of qubits $\{1,2\}$ and $\{3,4\}$:
\begin{eqnarray}
\left\{\ket{\Psi(q_i,q_j)}\right\}\simeq \mathcal{H}^S_\mathcal{G} \simeq 1\otimes 1 \oplus 0\otimes0 \nonumber\\
\subset \frac{1}{2}\otimes\frac{1}{2}\otimes\frac{1}{2}\otimes\frac{1}{2}\simeq\mathcal{H_G}\otimes\mathcal{H_G}.
\end{eqnarray}
Here, $1\otimes 1$ ($0\otimes0$) denotes the tensor product of the subspaces with total
spin-$1$($0$) for the qubits $\{1,3\}$ and $\{2,4\}$ (see Fig.\,\ref{4Qubits}b).
This will prove point i) of the conjecture. In proving this result, I will characterize the hidden topological order of the symmetrized states in Eq.\,(\ref{Symm States qq}). 
Furthermore, I will show that the topological charges of the symmetrized model are in one to one correspondence with the 4-qubit states in $\mathcal{H}^S_\mathcal{G}$ of the form:
\begin{eqnarray}
\ket{q_i\odot q_j} \propto 
\ket{q_i}\otimes\ket{q_j}+\ket{q_j}\otimes\ket{q_i}.
\end{eqnarray}
Finally, I will obtain the fusion and braiding rules characterizing the symmetrized model.

\subsection{Characterization of symmetrized states through spin-$1$ string operators}

\begin{figure}[t!]
\includegraphics[width=\linewidth]{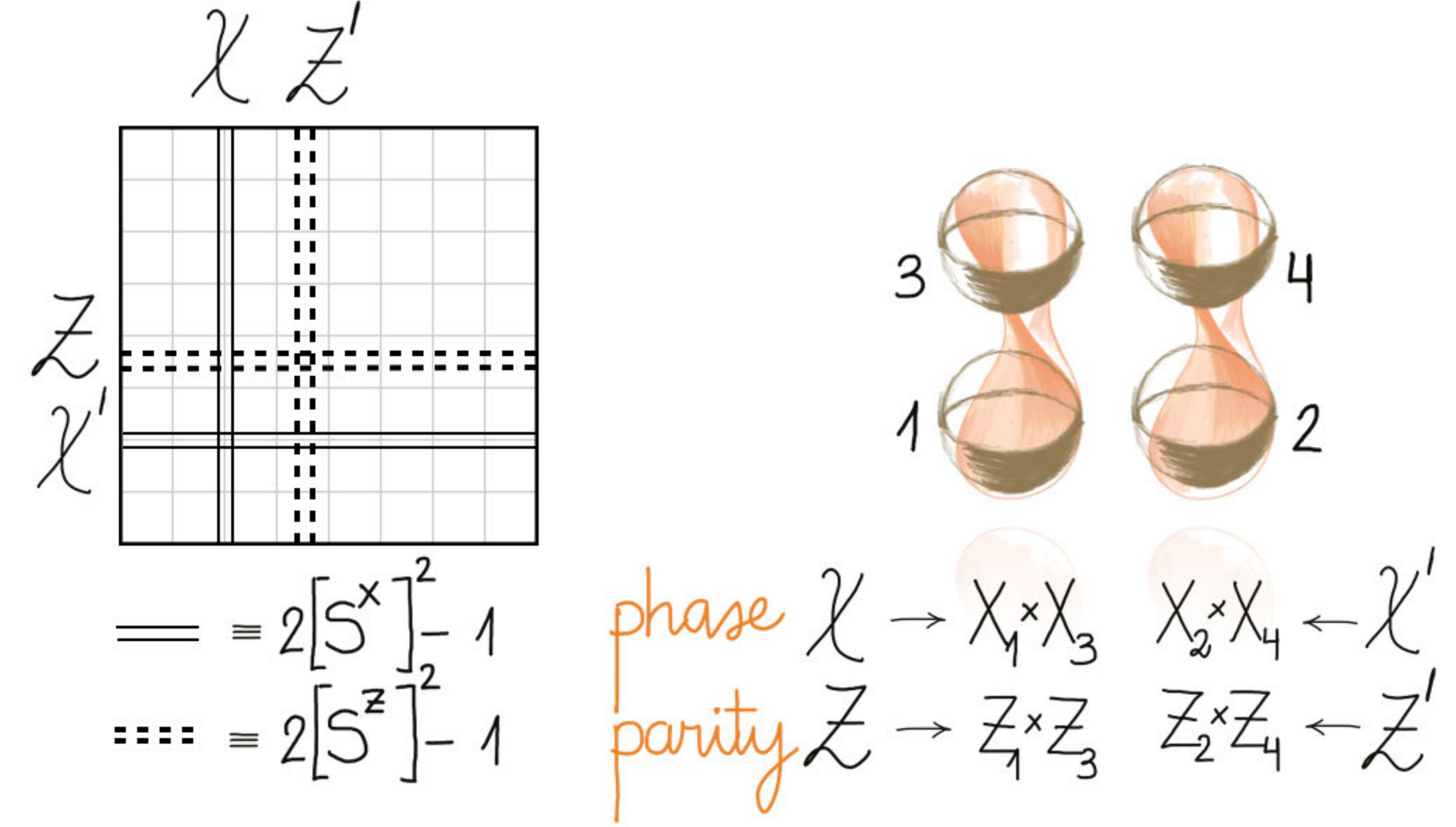}
\caption{\textbf{Spin-1 string operators}. The string operators $\mathcal{X},\mathcal{Z}$ ($\mathcal{X}',\mathcal{Z}'$) on the normal (dual) lattice completely characterize the manifold of symmetrized states on a torus. These operators correspond in the global symmetric subspace of 4-qubits to the phase and parity operators of the pairs of qubits $\{1,3\}$ and $\{2,4\}$. 
}
\label{StringOperators}
\end{figure}

Let me start by defining the following spin-$1$ string operators on the torus:
\begin{eqnarray}
\mathcal{X}(\mathcal{X}^{\prime})&=&
\prod_{\ell \in \mathcal{C}_{x}(\mathcal{C}_{y})} 
(2[S_{\ell}^x]^2-1)\nonumber\\
\mathcal{Z}^{\prime}(\mathcal{Z})&=&
\prod_{\ell \in \widetilde{\mathcal{C}}_{x} (\widetilde{\mathcal{C}}_{y})} 
(2[S_{\ell}^z]^2-1).
\end{eqnarray}
They trivially fulfill $\mathcal{X}^2(\mathcal{X}^{\prime 2})=\mathcal{Z}^2(\mathcal{Z}^{\prime 2})=\mathbbm{1}$, so that their eigenvalues are $\pm 1$.
We can write them as projections of tensor products of spin-$\frac{1}{2}$ string operators in each copy in the form:
\begin{eqnarray}
\mathcal{X}(\mathcal{X}^{\prime})\mathbf{P}=\mathbf{P}\hspace{-2pt}
\prod_{\ell \in \mathcal{C}_{x}(\mathcal{C}_{y})} \hspace{-1pt}
P_\ell^{} (\sigma^x_\ell \otimes \sigma^x_\ell)P_\ell
=\mathbf{P}(X_{1(2)} \otimes X_{3(4)})\nonumber\\
\mathcal{Z}^{\prime}(\mathcal{Z})\mathbf{P}=\mathbf{P}\hspace{-2pt}
\prod_{\ell \in \widetilde{\mathcal{C}}_{x} 
(\widetilde{\mathcal{C}}_{y})} \hspace{-1pt}
P_\ell^{} (\sigma^z_\ell \otimes \sigma^z_\ell)P_\ell
=\mathbf{P}(Z_{2(1)} \otimes Z_{4(3)}). \nonumber
\end{eqnarray}
From the above expressions, we see that the operators $\mathcal{Z},\mathcal{X}$ correspond, respectively, to the parity ($Z\otimes Z$) and phase ($X\otimes X$) of the pair of qubits $\{1,3\}$. Conversely, the operators $\mathcal{Z}^{\prime},\mathcal{X}^{\prime}$ correspond to the parity and phase operator of the pair of qubits $\{2,4\}$ (see Fig.\,\ref{StringOperators}). 

The symmetrized states in Eq.\,(\ref{Symm States qq}) are eigenstates of 
$\mathcal{X}$ and $\mathcal{Z}^{\prime}$:
\begin{eqnarray}
\mathcal{X}  \ket{\Psi(q_i,q_j)} =
\mathbf{P}\left(X_1\ket{\Phi_{q_i}}  \otimes X_3\ket{\Phi_{q_j}}\right)
=\pm1 \nonumber\\
\mathcal{Z}^{\prime}  \ket{\Psi(q_i,q_j)} =
\mathbf{P} \left(Z_2\ket{\Phi_{q_i}}  \otimes Z_4\ket{\Phi_{q_j}}\right)
=\pm1.
\end{eqnarray}
We can classify them according to the different eigenvalues of 
$\mathcal{X}$ and $\mathcal{Z}^{\prime}$ (see Table I).
Within each subspace the states are related to each other by the operators 
$\mathcal{X}^{\prime}$ and $\mathcal{Z}$.
For example, we have
\begin{eqnarray}
\ket{\Psi(e,e)}&=&\mathbf{P}(\ket{\Phi_e}\otimes \ket{\Phi_e})=\nonumber\\
&=&\mathbf{P}(X_2\otimes X_4)(\ket{\Phi_1}\otimes \ket{\Phi_1})\nonumber\\
&=&\mathcal{X}^\prime\mathbf{P}(\ket{\Phi_1}\otimes \ket{\Phi_1})=
\mathcal{X}^\prime\ket{\Psi(1,1)}.
\end{eqnarray}

\begin{table}[h]
\centering
\begin{tabular}{|r|r||lllllll|}
\hline
  $\mathcal{X}$ & $\mathcal{Z}'$ 
& \multicolumn{6}{c}{Symmetrized State} &\\ [0.5ex] 
\hline \hline

$1$ & $1$ & & & $\Psi(1,1)$  & $\xrightarrow{\mathcal{X}^\prime}$
& $\Psi(e,e)$ && \\
&     & & & &
$\xrightarrow{\mathcal{Z}^{\,\,}}$ & $\Psi(m,m)$ & $\xrightarrow{\mathcal{X}^\prime}$
&$\Psi(em,em)$ \\
\hline 
$1$ & $-1$ & & & $\Psi(1,e)$  &  $\xrightarrow{\mathcal{Z}^{\,\,}}$& $\Psi(m,em)$ &&\\
\hline
$-1$& $1$ & & &$\Psi(1,m)$  & $\xrightarrow{\mathcal{X}^\prime}$ & $\Psi(e,em)$ &&\\
\hline
$-1$& $-1$ & & &$\Psi(e,m)$  &$\xrightarrow{\mathcal{X}^\prime}$ & $\Psi(1,em)$ &&\\
\hline
\end{tabular}
\label{TableMicroscopic}
\caption{Symmetrized states are clasiffied by the eigenvalues of the spin-1 string operators $\mathcal{X}$ and $\mathcal{Z}'$. Within each subspace the states are connected through the operators $\mathcal{X}'$ and $\mathcal{Z}$.
}
\end{table}

\subsection{Correspondence between symmetrized many-body spin-$1$ states and 4-qubit states}

Let me consider a basis for the symmetric subspace of four qubits 
$\mathcal{H}^S_\mathcal{G}$ in the form:
\begin{eqnarray}
\ket{q_i \odot q_j}=\alpha_{ij}(\ket{q_i}\otimes\ket{q_j}+\ket{q_j}\otimes\ket{q_i}),
\label{SymCharges}
\end{eqnarray}
with $\alpha_{ij}^{-1}=\sqrt{1+\delta_{ij}}$.
These states are eigenstates of the operators $X_1\otimes X_3$ and $Z_2\otimes Z_4$ 
[46] and therefore have well defined phase and parity for the pairs of qubits \{1,3\} and \{2,4\}, respectively. They fulfill:
\begin{eqnarray}
&&X_1\otimes X_3\ket{q_i \odot q_j} = \nonumber\\
&&\alpha_{ij}X_1\otimes X_3 
\left[1+\mathsf{SWAP}_{13}\mathsf{SWAP}_{24}\right] \ket{q_j}\otimes \ket{q_i}= \nonumber \\
&&\alpha_{ij}\left[1+\mathsf{SWAP}_{13}\mathsf{SWAP}_{24}\right] X_1\ket{q_j}\otimes X_3\ket{q_i}= \pm 
\ket{q_i \odot q_j},\nonumber
\end{eqnarray}
and, similarly, for $Z_2 \otimes Z_4$.
Within each eigensubspace the states are connected to each other through the operators 
$X_2 \otimes X_4$ and $Z_1 \otimes Z_3$ as shown in Table II.

\begin{table}[h]
\label{table:Qubits}
\centering
\begin{tabular}{|rr||lllllll|}
\hline
$X_1X_3$ & $Z_2Z_4$ 
& \multicolumn{6}{c}{4-qubit Symmetric State}&\\ 
[0.8ex] 
\hline \hline
$1$ & $1$ & & & $1\odot 1$  & $\xrightarrow{X_2X_4}$ & $e\odot e$ &&\\
    &     & & & & $\xrightarrow{Z_1Z_3\,}$ & $m \odot m$ & $\xrightarrow{X_2X_4}$ 
    & $em \odot em$\\
\hline 
$1$ & $-1$ & & & $1 \odot e$  &  $\xrightarrow{Z_1Z_3\,}$& $m \odot em$ &&\\
\hline
$-1$& $1$ & & &$1 \odot m$  & $\xrightarrow{X_2X_4}$ & $e \odot em$ &&\\
\hline
$-1$& $-1$ & & &$e \odot m$  &$\xrightarrow{Z_1Z_3\,}$ & $1\odot em$ &&\\
\hline
\end{tabular}
\caption{Symmetric 4-qubit states are classified by the eigenvalues of the operators $X_1X_3$ and $Z_2Z_4$. Within each eigensubspace, the states are connected through the operators $X_2X_4$ and 
$Z_1Z_3$. }
\end{table}
\begin{figure}[t]
\includegraphics[width=\linewidth]{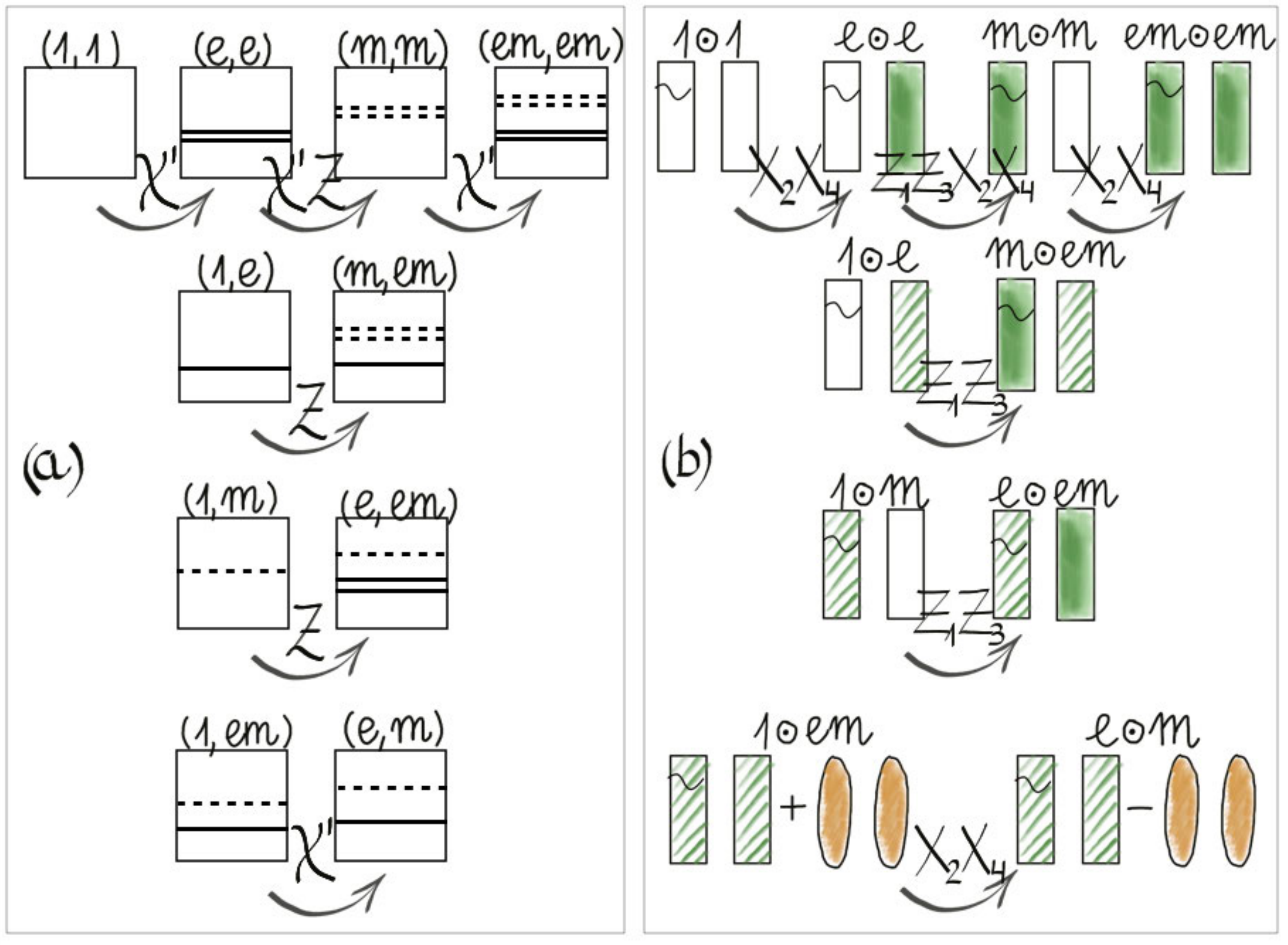}
\caption{\textbf{Microscopic vs global algebra}. The algebra of operators connecting the different many-body symmetrized states (\textbf{a}) is equivalent to the one of operators connecting the 4-qubit states in the symmetric subspace (\textbf{b}). The spin-1 string operators $\mathcal{X},\mathcal{X}^\prime,\mathcal{Z},\mathcal{Z}^\prime$ correspond, respectively, to the parity and phase operators for pairs of qubits $X_1X_3,X_2X_4,Z_1Z_2,Z_3Z_4$. The symmetric 4-qubit states $\ket{q_i \odot q_j}$ are represented in terms of products of triplets and singlets for pairs of qubits $\{1,3\}$ and $\{2,4\}$ (see inset in Fig.\,\ref{TopologicalCharges}).}
\label{Mapping}
\end{figure}	

If we establish the correspondence:
\begin{eqnarray}
\mathcal{X}(\mathcal{X}^\prime) &\longleftrightarrow& X_{1(2)}\otimes X_{3(4)}\nonumber \\ 
\mathcal{Z}(\mathcal{Z}^\prime) &\longleftrightarrow& Z_{1(2)}\otimes Z_{3(4)}, 
\end{eqnarray}
between string operators and parity and phase operators for pairs of qubits, the two tables 
I and II are equivalent (see Fig.\,\ref{Mapping}). We can therefore conclude that the ground state subspace of many-body symmetrized states is isomorphic to the symmetric subspace of four qubits, $\mathcal{H}^S_\mathcal{G}$. When establishing the one to one correspondence between many-body symmetrized states and 4-qubit states, we have to take into account the fact that the symmetrized states $\ket{\Psi(q_i,q_j)}$ within each eigensubspace (states within the same row in Table I) are not orthogonal to each other. For example, we have:
\begin{eqnarray}
\langle \Psi(1,1) \vert \Psi(e,e) \rangle=\langle \Psi(1,1)\vert \,\mathcal{X}'\, \vert 
\Psi(1,1)\rangle\ne0.
\end{eqnarray}
The states can be made orthogonal by making symmetric and antisymmetric superpositions of the form:
\begin{eqnarray}
\ket{\Psi(1,1)} \pm \ket{\Psi(e,e)}=(1\pm\mathcal{X}') \ket{\Psi(1,1)},
\end{eqnarray}
which correspond to different eigenvalues of the operator $\mathcal{X}'$. 
The actual isomorphism is:
\begin{table}[h]
\centering
\begin{tabular}{lllll}
$\eta_{\pm}(1\pm\mathcal{X}^\prime)$ & $\ket{\Psi(1,1)}$ &
$\longleftrightarrow$    $\frac{1}{\sqrt{2}}(1\pm X_2X_4)$ &$\ket{1\odot 1}$
\\
$\eta_{\pm}(1\pm\mathcal{Z})$ & $\ket{\Psi(1,1)}$ &
$\longleftrightarrow$    $\frac{1}{\sqrt{2}}(1\pm Z_1Z_3)$ &$\ket{1\odot 1}$
\\
$\eta_{\pm}(1\pm\mathcal{Z})$ & $\ket{\Psi(1,e)}$ &
$\longleftrightarrow$    $\frac{1}{\sqrt{2}}(1\pm Z_1Z_3)$ &$\ket{1\odot e}$  
\\
$\eta_{\pm}(1\pm\mathcal{X}^\prime)$ & $\ket{\Psi(1,m)}$ &
$\longleftrightarrow$  $\frac{1}{\sqrt{2}}(1\pm X_2X_4)$ &$\ket{1\odot m}$ 
\\
$\eta_{\pm}(1\pm\mathcal{X}^\prime)$ & $\ket{\Psi(e,m)}$ &
$\longleftrightarrow$   $\frac{1}{\sqrt{2}}(1\pm X_2X_4) $ &$\ket{e \odot m}.$ 
\end{tabular}
\end{table}



\section{CHARGES, FUSION RULES AND S-MATRIX}

\subsection{The charges of the model}

The charges $\ket{\mathcal{Q}_{ij}}$ of the symmetrized model are given by the superpositions of the symmetrized many-body states 
$\ket{\psi(q_i,q_j)}$ in Eq.\,(\ref{Symm States qq}) that satisfy the condition of mimimum entanglement entropy. 
Before symmetrizing, the charges of two copies of the toric code model correspond in the effective subspace of 4-qubits to the tensor product states $\ket{q_i}\otimes \ket{q_j}$ defined in 
Eq.\,(\ref{ProductStates}).
Therefore, for two-copies of the toric code model, to satisfy the condition of minimum entanglement entropy is equivalent in the subspace of 4-qubits to guarantee that  the pairs $\{1,3\}$ and $\{2,4\}$ have well defined $x$- and $z$- components, respectively.

Let me assume here the following intuitive result: Since the subspace of symmetrized many-body states is isomorphic to the symmetric subspace of 4-qubits $\mathcal{H}_\mathcal{G}^S$, finding the basis of minimum entanglement entropy states (MES's) for the symmetrized model is equivalent to finding the basis for 4-qubits 
that better defines the $x$- and $z$-components for qubits $\{1,3\}$ and $\{2,4\}$ within $\mathcal{H}_\mathcal{G}^S$. This result can indeed be proved to be true. By considering a partition on the torus, the microscopic basis of MES's of the symmetrized model can be 
obtained \cite{Sebastian}. This calculation is similar to the one performed to obtain the charges for a single copy of the toric code model \cite{Vishwanath1,Vishwanath2}. This proof will be presented elsewhere \cite{Sebastian}. 

It is straightforward to see that the basis in the symmetric subspace $\mathcal{H}_\mathcal{G}^S$ that allows for a better definition of the $x$- and $z$-components of qubits 
$\{1,3\}$ and $\{2,4\}$ is precisely the basis $\ket{q_i \odot q_j}$ introduced above in Eq.\,(\ref{SymCharges}) (see Appendix). Therefore we have that:
\begin{equation}
\ket{\mathcal{Q}_{ij}}=\ket{q_i \odot q_j},
\end{equation}
and the charges of the model are:
\begin{eqnarray}
\begin{array}{ccc}
\mathcal{I}= 1\odot 1 &  \mathcal{E}=1\odot e &  \mathcal{E}_2=e\odot e\\
\mathcal{M}=1\odot m &  (\mathcal{EM})_1, (\mathcal{EM})_2&  \mathcal{E}_2\mathcal{M}=e\odot em\\
\mathcal{M}_2=m\odot m &  \mathcal{E}\mathcal{M}_2=em\odot m &  
\mathcal{E}_2\mathcal{M}_2=em\odot em,
\end{array}
\end{eqnarray}
with $(\mathcal{EM})_1=e\odot m$ and $(\mathcal{EM})_2=em\odot 1$.

\begin{figure}[t]
\includegraphics[width=\linewidth]{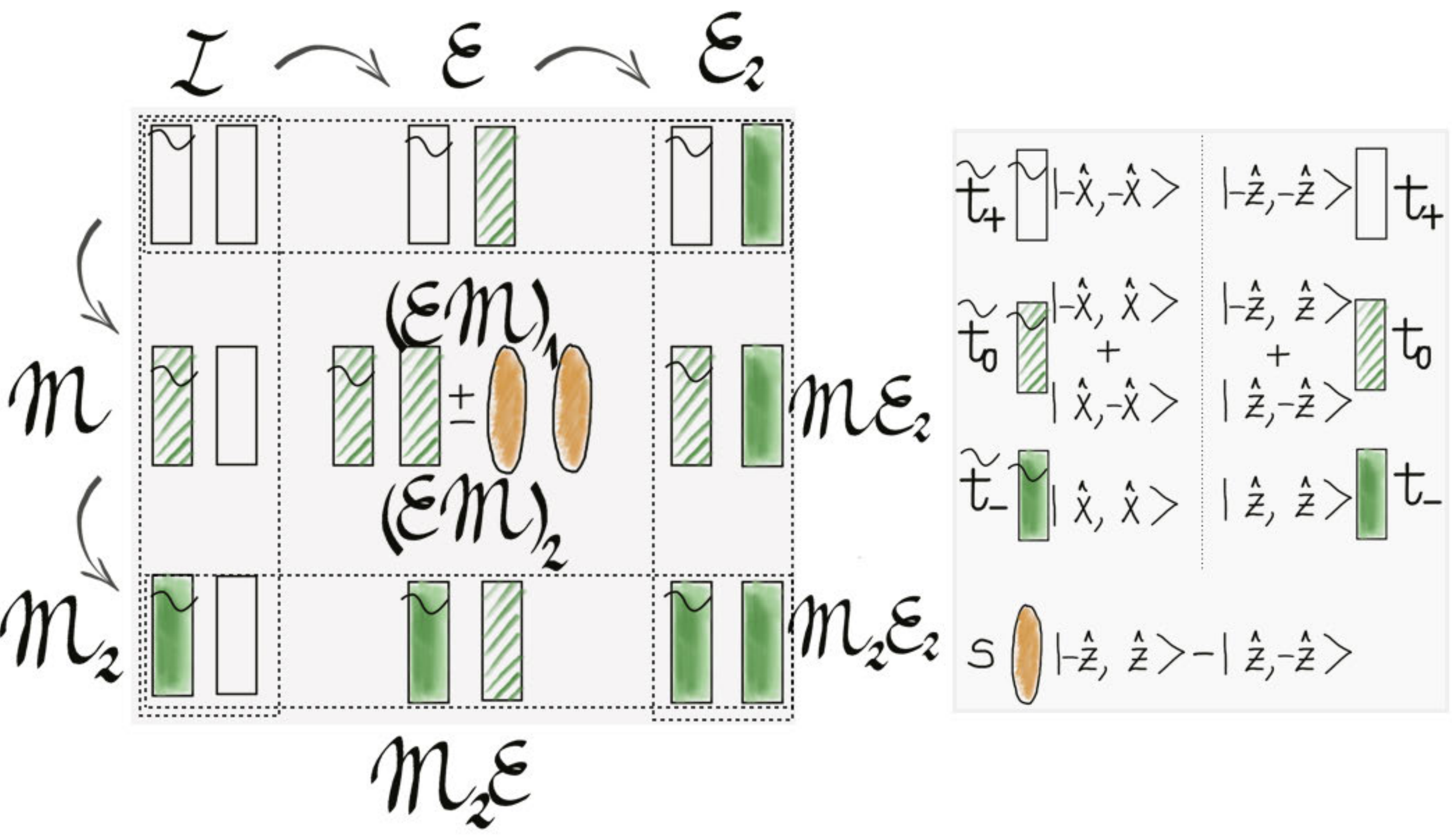}
\caption{\textbf{Topological charges of the symmetrized model}. The ten topological charges of the model are represented in terms of triplet and singlet states for the pairs of qubits $\{1,3\}$ and $\{2,4\}$. The trios $\{\mathcal{I},\mathcal{E},\mathcal{E}_2\}$, $\{\mathcal{I},\mathcal{M},\mathcal{M}_2\}$ fulfill a spin-1 algebra and behave as copies of an Ising model.
}
\label{TopologicalCharges}
\end{figure}	

It is useful to write the states $\ket{q_i \odot q_j}$ in terms of tensor products of triplets and singlets  for the pairs of qubits $\{1,3\}$ and $\{2,4\}$.
If we define 
\begin{eqnarray}
	\begin{array}{ll}
\ket{\widetilde{t}_+}=\ket{\hat{x},\hat{x}} 
& \ket{t_+}=\ket{\hat{z},\hat{z}}\\
\ket{\widetilde{t}_0}= \frac{1}{\sqrt{2}}(\ket{\hat{x},-\hat{x}}+\ket{-\hat{x},\hat{x}}) ,
& \ket{t_0}=\frac{1}{\sqrt{2}}(\ket{\hat{z},-\hat{z}}+\ket{-\hat{z},\hat{z}}) \nonumber\\
\ket{\widetilde{t}_-} =\ket{-\hat{x},-\hat{x}} & 
\ket{t_-}=\ket{-\hat{z},-\hat{z}}\\
\hline
\ket{s}=\frac{1}{\sqrt{2}}(\ket{\hat{z},-\hat{z}}-\ket{-\hat{z},\hat{z}}),
\end{array}
\end{eqnarray}
most of the states $\ket{q_i\odot q_j}$ can be written as products of triplets 
$\ket{\widetilde{t}_\alpha} \otimes \ket{t_\beta}$ (see Fig.\,\ref{Mapping}). The states $\ket{e\odot m}$ and $\ket{em \odot1}$ are, respectively, symmetric and antisymmetric superpositions of $\ket{\widetilde{t}_0} \otimes \ket{t_0}$ and $\ket{s}\otimes\ket{s}$.

\subsection{The fusion rules}

The fusion rules \cite{Nayak2008,Preskill} follow from the algebra of spin operators in the symmetrized subspace $\mathcal{H}^S_\mathcal{G}$. On the one hand, we have that  $\{\mathcal{I},\mathcal{E},\mathcal{E}_2\}$ and $\{\mathcal{I},\mathcal{M},\mathcal{M}_2\}$ fulfill a spin-1 algebra:
\begin{eqnarray}
&&\mathcal{I} \xrightarrow{\frac{X_2+X_4}{\sqrt{2}}\left(\frac{Z_1+Z_3}{\sqrt{2}}\right)} 
\mathcal{E} (\mathcal{M}) \longrightarrow 
\mathcal{E}_2 (\mathcal{M}_2)\nonumber\\
&&\mathcal{I} \xrightarrow{\hspace{0.6cm}X_2X_4(Z_1Z_3)\hspace{1.63cm}} \mathcal{E}_2(\mathcal{M}_2),
\end{eqnarray}
with $\frac{X_2+X_4}{2}$($\frac{Z_1+Z_3}{2}$) corresponding to the $x$-component of an effective global spin-1 operator $S^x_\mathcal{G}$, and
$X_2X_4$($Z_1Z_3$) corresponding to $1-2[S^x_\mathcal{G}]^2$.
It therefore follows that:
\begin{eqnarray}
\begin{array}{ccccc}
\mathcal{E} \times \mathcal{E}= \mathcal{I}+ \mathcal{E}_2 & &
\mathcal{E} \times \mathcal{E}_2=\mathcal{E} & &
\mathcal{E}_2 \times \mathcal{E}_2=\mathcal{I} \nonumber\\
\mathcal{M} \times \mathcal{M}=\mathcal{I}+ \mathcal{M}_2 & &
\mathcal{M} \times \mathcal{M}_2=\mathcal{M} & &
\mathcal{M}_2 \times \mathcal{M}_2=\mathcal{I}. \\
\end{array}
\end{eqnarray}
\newline
On the other hand, we have that 
\begin{eqnarray}
&&\mathcal{I} \xrightarrow{\frac{Z_1X_4+Z_3X_2}{\sqrt{2}}} (\mathcal{EM})_1\nonumber\\
&&\mathcal{I} \xrightarrow{\frac{Z_3X_4+Z_1X_2}{\sqrt{2}}} (\mathcal{EM})_2,
\end{eqnarray}
so that 
\begin{eqnarray}
\begin{array}{cc}
\mathcal{E} \times \mathcal{M}= (\mathcal{EM})_1+ (\mathcal{EM})_2 \nonumber \\
(\mathcal{EM})_1 \times (\mathcal{EM})_1=
(\mathcal{EM})_2\times (\mathcal{EM})_2=1+\mathcal{E}_2 \mathcal{M}_2\nonumber \\
(\mathcal{EM})_1\times (\mathcal{EM})_2=\mathcal{E}_2 +\mathcal{M}_2. \\
\end{array}
\end{eqnarray}

The other fusion rules follow from these non-trivial ones by using the associative property. The subsets of charges $\{\mathcal{I},\mathcal{E},\mathcal{E}_2\}$ and $\{\mathcal{I},\mathcal{M},\mathcal{M}_2\}$ behave internally like two type of Ising models \cite{Preskill,Nayak2008}, 
with $\mathcal{E},\mathcal{M}$ being two types of Ising quasiparticles, and $\mathcal{E}_2, \mathcal{M}_2$ the corresponding two types of fermions. But, interestingly, the charges $\mathcal{E}$ and $\mathcal{M}$ are also non-Abelian with respect to each other, having two different fusion channels,  $(\mathcal{EM})_1$ or $(\mathcal{EM})_2$. 
We see how, through symmetrization, the Abelian charges are converted into non-Abelian ones with non-trivial fusion channels.


\subsection{The S-matrix}

The $S$-matrix of the model is easily obtained as the modular transformation relating the two basis of charges $\mathcal{Q}$ and 
$\widetilde{\mathcal{Q}}$ corresponding to the -$x$ and -$y$ direction of the torus \cite{Vishwanath1,Vishwanath2,Vishwanath3}. For the model I discuss here, we have:
\begin{eqnarray}
&&\mathcal{S}_{ij,i'j'}=
\langle q_i \odot q_j \vert \widetilde{q_{i'}\odot q_{j'}} \rangle= \nonumber \\
&&\langle q_i \odot q_j \vert \mathsf{SWAP}_{12} \mathsf{SWAP}_{34} 
\vert  q_{i'}\odot q_{j'} \rangle =  \\
&&\langle i \odot j \vert \mathsf{H}_1  \mathsf{H}_3 \cdot
\mathsf{SWAP}_{12} \mathsf{SWAP}_{34} \cdot
\mathsf{H}_1  \mathsf{H}_3 \vert i' \odot j' \rangle,\nonumber
\end{eqnarray}
where I have used that $\ket{q_i \odot q_j}=\mathsf{H}_1 \otimes \mathsf{H}_3 \ket{i\odot j}$, with $\ket{i\odot j}\propto \ket{i} \odot \ket{j}+\ket{j}\odot \ket{i}$. Therefore, we obtain that:
\begin{eqnarray}
\mathcal{S}_{ij,i'j'}&=&
\langle i \odot j \vert 
\mathsf{SWAP}_{12} \mathsf{SWAP}_{34} \cdot \mathsf{H}_2  \mathsf{H}_4
\mathsf{H}_1  \mathsf{H}_3 \vert i' \odot j' \rangle=\nonumber\\
&&\langle i \odot j \vert S \otimes S \vert i' \odot j' \rangle.
\label{SMatrixSym}
\end{eqnarray}
Thus the $S$-matrix of the symmetrized model is the projection onto the symmetric subspace of the tensor product of the $S$-matrices of the copies. It is illuminating to write the $S$-matrix in the basis of triplets and singlets $\{\ket{t_\alpha}\otimes\ket{t_\beta}, \ket{s}\otimes\ket{s}\}$.
In this basis, the matrix is block diagonal in the subspaces $1\otimes 1$ and $0\otimes 0$. Taking into account that 
\begin{eqnarray}
\bra{t_\alpha} \mathsf{H}\otimes \mathsf{H} 
\ket{t_\beta}= [e^{i \frac{\pi}{4}S^y}]_{\alpha,\beta},
\end{eqnarray}
with $S^y$ the $y$-component of a spin-1 operator,
the $S$-matrix takes the form:
\begin{eqnarray}
\mathcal{S}^\prime&=&
\mathsf{SWAP}_{tt}\,e^{i \frac{\pi}{4}S_y}\otimes e^{i \frac{\pi}{4}S_y}\oplus\mathbbm{1}_{ss}.
\label{SMatrixTriplet}
\end{eqnarray}
Here, $\mathsf{SWAP}_{tt}$ is a unitariy operator in the $1\otimes 1$ subspace that exchanges the two triplets and maps $\ket{t_\alpha}\otimes\ket{t_\beta}$ onto $\ket{t_\beta}\otimes\ket{t_\alpha}$, reordering the elements of the basis. Up to this reordering, within the subspace of triplets the $S$-matrix is the tensor product of two copies of the matrix
\begin{eqnarray}
e^{i \frac{\pi}{4}S_y}=
\left(
\begin{array}{ccc}
1 & -\sqrt{2} & 1 \\
-\sqrt{2} & 0 & \sqrt{2} \\
1 &\sqrt{2} & 1
\end{array}
\right).\label{SMatrixIsing}
\end{eqnarray}
The matrix (\ref{SMatrixIsing}) is precisely the $S$-matrix of the Ising (non-Abelian) model \cite{Preskill}. The $S$-matrix of the symetrized model in Eq.\,(\ref{SMatrixSym}) is obtained as $U^\dagger \mathcal{S}^\prime U$, where $U$ is the unitary transforming the basis of triplets and singlets $\{\ket{t_\alpha}\otimes\ket{t_\beta}, \ket{s}\otimes\ket{s}\}$ into the basis of states 
$\ket{i\odot j}$. The latter are given by $\ket{t_\alpha}\otimes\ket{t_\beta}$, for $(\alpha,\beta)\ne(0,0)$, and $\frac{1}{\sqrt{2}}(\ket{t_0}\otimes\ket{t_0}\pm\ket{s}\otimes\ket{s})$. This change of basis destroys the tensor product structure of $S^\prime$ in Eq.\,(\ref{SMatrixTriplet}) and couples the two Ising models.

\section{CONCLUSIONS AND OUTLOOK}

I have proposed a physical mechanism for the emergence of non-Abelian topological order from a set of physical microscopic degrees of freedom. Within my proposal non-Abelian order arises from the organization of a quantum many-body system in indistinguishable copies of the same collective state. The intuition behind this idea is simple. The freedom to assign quasiparticles to the identical copies gives origin to a set of degenerate states that are locally indistinguishable and are connected to each other by braiding of quasiparticles. This opens the path for non-Abelian braiding statistics.

I have presented a conjecture on topological models that arise from symmetrization of identical copies of an Abelian model. The manifold of many-body symmetrized states on a torus is proposed to be isomorphic to an effective symmetric global Hilbert space. This global space is spanned by symmetrizations of tensor products of Abelian topological charges, which define the quasiparticles of the symmetrized model. 
Similarly, the modular $S$-matrix is proposed to be obtained as the projection onto the symmetric global subspace of the tensor product of copies of the Abelian $S$-matrix. 

To illustrate the theory, I have analyzed the case of two copies of the toric code model. By defining appropriate spin-1 string operators, an isomorphism has been established between the space of microscopic symmetrized many-body ground states and a symmetric global subspace of four qubits. Using this correspondence, I have argued that the topological charges of the model satisfy the general conjecture and are obtained as symmetrization of products of the toric code quasiparticles. The symmetrized model has been shown to be non-Abelian, corresponding to two copies of an Ising model that are linked together in a non-trivial manner. It is interesting to investigate whether this non-Abelian model might be universal for quantum computation \cite{Nayak2008}.
The symmetrized topological charges can be shown to be the states of minimum entanglement entropy by considering a bipartition on the torus. The details of this result will be presented in an upcoming work \cite{Sebastian}. 

The ideas developed here for the case of two copies of the toric code model can be generalized to prove the conjecture in the general case. Starting with many-body operators characterizing the topological order of the Abelian copies, one can construct appropriate string operators for the symmetrized many-body states. The analysis of the algebra of such operators would allow one to identify the isomorphic symmetric global effective subspace. The topological charges would be then obtained through symmetrization of Abelian quasiparticles in such global subspace.

This formalism opens a path for the generation of non-Abelian models from known Abelian ones. It is challenging to investigate what type of non-Abelian models arise as symmetrization of Abelian copies. For instance, interesting many-body states and models might emerge from the symmetrization of copies of {\em chiral spin liquids}, {\em resonating valence bond states} or {\em topological insulators}. 
Conversely, it is intriguing to explore whether known non-Abelian models could be deconstructed into copies of Abelian ones. This is the case for the seminal example of a non-Abelian state, the Pfaffian state, which is expected to occur in fractional quantum Hall systems. This state can be written as two copies of a Laughlin (Abelian) state \cite{Paredes2012b}. I believe that similar constructions are also possible for other known instances of non-Abelian anyons and states,  such as Fibonacci anyons or $p$-wave superconductors. 

Finally, the investigation of the parent Hamiltonians associated to these symmetrized non-Abelian models is of crucial importance. My previous work for the case of two copies of the toric code model \cite{Paredes2012b} suggests that the interactions behind this type of order are local and involve a small number of spins. It is challenging to explore whether we can develop a recipe for the construction of non-Abelian parent Hamiltonians based on the Hamiltonians of the Abelian copies. This approach would help us deepen our understanding of the microscopic interactions underlying non-Abelian anyons, serving as a guide for their experimental realization \cite{Zoller}.

\section{ACKNOWLEDGMENTS}
I would like to thank Sebasti\'{a}n Montes, Guifr\'{e} Vidal and Germ\'{a}n Sierra 
for very useful discussions.

\section{Appendix: Basis of symmetrized topological charges}

I aim to find the basis of states of four qubits that better defines the individual $x$-components of spins $\{1,3\}$ and the $z$-components of spins $\{2,4\}$, within the subspace of states that is symmetric under the simultaneous exchange of spins $1$ and $3$, and $2$ and $4$.
Since the projection onto this symmetric subspace commutes with the total spin components $(X_1+X_3)/2$ and $(Z_2+Z_4)/2$, the elements of such basis must be eigenstates of these operators. These eigenstates are precisely $\{\ket{\widetilde{t_\alpha}}\otimes\ket{t_\beta}, \ket{s}\otimes\ket{s}\}$, with the triplets $\ket{\widetilde{t_\alpha}}$, $\ket{t_\beta}$, and the singlet  $\ket{s}$ defined above (see Fig.\,\ref{TopologicalCharges}). They are characterized by different eigenvalues of the operators $(X_1+X_3)/2$ and $(Z_2+Z_4)/2$, except for the states $\ket{\widetilde{t_0}}\otimes \ket{t_0}$ and $\ket{s}\otimes\ket{s}$, for which both spin components are zero. The symmetric and antisymmetric superpositions of these two states are the ones that better define the individual $x$-components of spins $\{1,3\}$ and the $z$-components of spins $\{2,4\}$, since such superpositions have vanishing expectation value for the operators $(Z_1+Z_3)/2$ and $(X_2+X_4)/2$. Therefore the basis that we are looking for is:
\begin{eqnarray}
\ket{\widetilde{t_\alpha}}\otimes\ket{t_\beta}, \,\, (\alpha, \beta) \ne (0,0) \nonumber \\
\frac{1}{\sqrt{2}}(\ket{\widetilde{t_0}}\otimes\ket{t_0}\pm\ket{s}\otimes\ket{s}).
\end{eqnarray}
This basis coincides with the one of symmetrized four qubit states $\ket{q_i\odot q_j}$ defined in Eq.\,(\ref{SymCharges}).

\end{document}